# Enhanced Seismicity Monitoring in the Rapid Scientific Response to the 2025 Santorini Crisis

Margarita Segou*[1], Foteini Dervisi[1,2], Xing Tan[3], Rajat Choudhary[1], Patricia Martínez-Garzón[4,5], Francesco Scotto di Uccio[6], Gregory Beroza[3], Genny Giacomuzzi[7], Claudio Chiarabba[7], Wayne Shelley[1], Stephanie Prejean[8], Jeremy Pesicek[9], John J. Wellik[9], Marco Bohnhoff[4], David Pyle[10], Costas Synolakis[11,12], Tom Parsons[13], Athanassios Ganas[14], William Ellsworth[3], Brian Baptie[1], Gaetano Festa[6,7], Piero Poli[15], and Warner Marzocchi[16]

[1] The Lyell Centre, British Geological Survey, Edinburgh, UK.
[2] School of GeoSciences, The University of Edinburgh, Edinburgh, UK.
[3] Department of Geophysics, Stanford University, Stanford, CA, USA.
[4] GFZ Helmholtz Centre for Geosciences, Potsdam, Germany.
[5] RWTH Aachen University, Aachen, Germany.
[6] Dipartimento di Fisica Ettore Pancini, Università degli Studi di Napoli Federico II, Naples, Italy.
[7] Istituto Nazionale di Geofisica e Vulcanologia (INGV), Rome, Italy.
[8] U.S. Geological Survey, Volcano Disaster Assistance Program, Anchorage, AK, USA.
[9] U.S. Geological Survey, Volcano Disaster Assistance Program, Vancouver, WA, USA.
[10] Department of Earth Sciences, University of Oxford, Oxford, UK.
[11] Department of Civil and Environmental Engineering, University of Southern California, Los Angeles, CA, USA.
[12] Division of Natural Sciences, Academy of Athens, Athens, Greece.
[13] U.S. Geological Survey, Pacific Coastal and Marine Science Center, Moffett Field, CA, USA.
[14] Institute of Geodynamics, National Observatory of Athens, Athens, Greece.
[15] Dipartimento di Geoscienze, Università degli Studi di Padova, Padua, Italy.
[16] Dipartimento di Scienze della Terra, dell' Ambiente e delle Risorse, Università degli Studi di Napoli Federico II, Naples, Italy.

*msegou@bgs.ac.uk [corresponding author]

*This version of the preprint has been officially approved by the USGS. It replaces the earlier version of the preprint, which was uploaded prior to the USGS internal review and was therefore not approved by the USGS.*

ORCiD (Margarita Segou): 0000.0001.8119.4019

ORCiD (Foteini Dervisi): 0000.0002.6257.3707

ORCiD (Xing Tan): 0009.0001.0919.3617

ORCiD (Rajat Choudhary): 0009.0006.0693.0219

ORCiD (Patricia Martínez-Garzón): 0000.0003.4649.0386

ORCiD (Francesco Scotto di Uccio): 0000.0001.6429.5600

ORCiD (Gregory Beroza): 0000.0002.8667.1838

ORCiD (Genny Giacomuzzi): 0009.0008.3135.4668

ORCiD (Claudio Chiarabba): 0000.0002.8111.3466

ORCiD (Wayne Shelley): 0000.0003.1879.1851

ORCiD (Stephanie Prejean): 0000.0003.0510.1989

ORCiD (Jeremy Pesicek): 0000.0001.7964.5845

ORCiD (John J. Wellik): 0000.0002.8099.5794

ORCiD (Marco Bohnhoff): 0000.0001.7383.635X

ORCiD (David Pyle): 0000.0002.2663.9940

ORCiD (Costas Synolakis): 0000.0003.0140.5379

ORCiD (Tom Parsons): 0000.0002.0582.4338

ORCiD (Athanassios Ganas): 0000.0002.1937.3283

ORCiD (William Ellsworth): 0000.0001.8378.4979

ORCiD (Brian Baptie): 0000.0001.6748.1740

ORCiD (Gaetano Festa): 0000-0002-2588-8160

ORCiD (Piero Poli): 0000.0002.6493.5142

ORCiD (Warner Marzocchi): 0000.0002.9114.1516

## Abstract

We used a deep learning workflow to enhance earthquake detection during the 2025 seismic unrest between Santorini and Amorgos islands to track the evolution of the unrest in near real-time. We analysed the continuous seismic waveforms daily (January 2– March 3, 2025) as the crisis unfolded. Our analysis enhanced the earthquake catalogue from around 4,000 to 80,000 earthquakes. The enhanced catalogue allowed this international expert group to identify the volcanic-tectonic character, clearly revealing burst-like, spasmodic seismicity swarms, which is a pattern associated with fluid-driven processes from early stages of the crisis. Detailed moment tensor inversions in early events characterised by a significant non-double couple component indicated the involvement of magmatic or high-pressure hydrothermal fluids driving the unrest. Concurrent Deep Learning-enhanced tomography efforts identified a third, deep magmatic reservoir beneath Anydros islet, consistent with pressure-driven processes. As of May 2026, volcanic-tectonic swarms with >200 earthquakes of $M_L$ > 4 occurring within only a few weeks, largely within episodic bursts of seismicity, have not been observed elsewhere.

## Second-language abstract

Η χρήση τεχνητής νοημοσύνης για την παρακολούθηση της πρόσφατης σεισμικής κρίσης στη Σαντορίνη σε σχεδόν πραγματικό χρόνο μας φέρνει μπροστά σε μία νέα πραγματικότητα στην αντιμετώπιση φυσικών καταστροφών. Στο πλαίσιο της παρούσας έρευνας, διεθνής ομάδα σεισμολόγων ανέλυε σε καθημερινή βάση τις συνεχείς σεισμικές κυματομορφές κατά τη διάρκεια της κρίσης. Ο κατάλογος σεισμών που προέκυψε με τη βοήθεια αλγορίθμων τεχνητής νοημοσύνης φθάνει τους 80.000 σεισμούς, αριθμός πολύ μεγαλύτερος των 4.000 σεισμών που εντοπίστηκαν με συμβατικές μεθόδους. Ο εμπλουτισμένος αυτός κατάλογος τεχνητής νοημοσύνης αποκάλυψε από τις πρώτες ημέρες της κρίσης το σύνθετο ηφαιστειακό-τεκτονικό χαρακτήρα της περιοχής, με έντονους παλμούς σεισμικότητας που υποδηλώνουν διεργασίες μετακίνησης ρευστών. Επιπλέον, πλήθος μηχανισμών γένεσης, που χαρακτηρίζουν τις σεισμικές διαρρήξεις, υποδεικνύουν την ύπαρξη μαγματικών ή υδροθερμικών διαδικασιών. Η εφαρμογή σεισμικής τομογραφίας με βάση τον εμπλουτισμένο κατάλογο αποκαλύπτει την ύπαρξη ενός μαγματικού θαλάμου κάτω από την

Άνυδρο, εύρημα που συνάδει με την εμφάνιση σεισμικότητας σε συνθήκες υπερπίεσης. Έως σήμερα , σε παγκόσμιο επίπεδο, δεν έχουν παρατηρηθεί πλησίον ηφαιστείων αντίστοιχες κρίσεις με >200 σεισμούς μεγέθους $M_L > 4$ που να συμβαίνουν κυρίως εντός επεισοδιακών εξάρσεων σεισμικότητας μέσα σε μόλις λίγες εβδομάδες.

## Introduction

Santorini is a mature and active arc volcano with at least 12 Plinian eruptions in the last 360 kyr, including the catastrophic Minoan eruption 3,600 years ago. Four of these eruptions involved caldera collapses (T. Druitt et al., 1999, 2024). The most recent large explosive eruption of Santorini was a submarine pumice eruption in 726 CE. Since then, eruptions have involved extrusion of lava and mild ash explosions on the intra-caldera Kameni islets (Preine et al., 2024; Pyle and Elliott, 2006). The last eruption was in 1950 (Georgalas, 1953).

Santorini is part of the trans-tensional “Christiana – Santorini – Kolumbo” rift zone that forms the extinct Christiana centre in the southwest through the Kolumbo submarine volcano and submarine cone field to the northeast (Fig. 1; Preine et al., 2022) and beyond towards Amorgos. This central part of the Hellenic Volcanic Arc with its complex and active tectonics has been termed the Santorini – Amorgos zone of crustal weakness (Bohnhoff et al., 2006). Not all activity in this zone is volcanic. The region experienced the 1956 earthquake doublet, when a magnitude ~7.5 event—likely on the Amorgos Fault—was followed by a 6.0 triggered earthquake (Brüstle et al., 2014) at greater hypocentral depth closer to Santorini (Brüstle et al., 2014; Kalligeris et al., 2025; Leclerc et al., 2024; Okal et al., 2009). The 1956 M7.5 event was the largest earthquake in the entire Eastern Mediterranean area in more than a century and caused violent shaking with severe damage in Amorgos and Santorini. The 1956 event also generated a tsunami with 5 – 20 m run-up in the near field, and >1 m within a distance of 100 km (Kalligeris et al., 2025).

Following the 1950 eruption, the Santorini volcanic complex remained largely quiescent for many years. Between January 2011 and March 2012, there was an extended episode of seismic unrest and inflation, with up to 15 cm of uplift of a part of the Kameni islands. This was interpreted as the intrusion of two pulses of magma into a shallow reservoir at around 4 km depth beneath the northern part of the caldera (Parks et al., 2015). Seismicity reached extremely shallow depths (to around 1 km), but no eruption ensued (Newman et al., 2012; Parks et al., 2012, 2015).

The broader volcanic field of Santorini, beyond the main caldera, has been seismically active within the modern instrumental era. Newman et al. (2012) suggested that in the 60 years before the 2011 unrest, microseismicity was focused on the submarine Kolumbo volcano that last erupted in 1650. Bohnhoff et al. (2006) used island-based temporary microseismic monitoring to observe localised microseismic clustering immediately below Kolumbo volcano, possibly representing its magma chamber. Within the latter study, seismicity further to the NE around Anydros islet was interpreted to reflect localised activity that might be diagnostic of local pathways of upward migrating fluids within an overall zone of crustal weakness, or alternatively to represent developing volcanic activity.

Today, Santorini is densely populated with more than 15,000 permanent residents (Hellenic Statistical Authority, 2023) and is a major cultural and economic hub hosting over than 3 million tourists annually by air, ferry and cruise ships (Institute of Greek Tourism Confederation, 2026). From the onset of the seismic crisis in early 2025, it quickly became evident that monitoring efforts faced many challenges, including high cultural noise at local land stations, as well as tracking the offshore seismic sequence from neighbouring islands. The seismic unrest remained intense, with five events $M_L > 5.0$ offshore Santorini within the first two weeks of February 2025, and the seismic evolution was poorly constrained. As a consequence of the unrest, the Civil Protection of Greece declared a state of emergency between February 6 and March 3, 2025, which caused concern among residents and triggered unsanctioned evacuations, partially disrupting the tourist season of Santorini, a contributor of ~2.5% of the annual Gross Domestic Product of Greece. In this context, high-resolution, near real-time monitoring and the timely interpretation of the early indicators of volcanic-seismic unrest is essential for effective emergency response to potential volcanic, seismic, and tsunami hazards (Fig. 1). Our expert group relayed results as these became available to Prof. Costas Synolakis of the Academy of Athens, who was participating in a relevant national committee on the Santorini unrest, and was working with colleagues (Kalligeris et al., 2025) on modelling tsunami generation from possible local tectonic and volcanic sources, a real concern given the history of the volcano. Prof. Costas Synolakis became the liaison between our expert group and the committee.

The integration of deep learning in seismic monitoring workflows enables rapid and comprehensive real-time analysis of seismic activity. This study recounts how deep learning (DL) tools were used to accelerate the pace of delivery of information content from computationally demanding scientific products and make them available in time to inform scientific and policy response during volcanic crises. DL tools complemented and extended other more traditional resources, including digital helicorders and spectrograms, which also provided essential information as the crisis unfolded. While an overview of methods is presented in the Supplementary Material (refer to Overview of Methods Section), this manuscript is organised chronologically, following the progression of the scientific analysis, using information *as it became available* (i.e., without retrospective analysis and interpretation) during February and March 2025. This real-time perspective describes how the outstanding questions at that time were addressed, including: What are the characteristics of the seismic crisis? Is it tectonically or volcanically driven? Are the known active faults engaged in the evolving phenomenon? Does all this information indicate an increased likelihood of an eruption?

## February 1: Improving the Earthquake Catalogue

Faced with the increased earthquake rates, an international team of scientists collaborated remotely to develop information that could be used to inform the scientific and societal response of the Santorini unrest. The first of these efforts sought to improve the earthquake catalogue in terms of detection threshold and event locations. It is now widely accepted that DL earthquake detection can offer up to 10 times the number of seismic events (e.g., Tan et al., 2021) compared to traditional techniques that rely on automatic picking and use short-term/long-term amplitude ratios, such as STA – LTA (e.g., Allen, 1978) or other characteristic-function-based methods that are less effective in detecting small magnitude or overlapping events, especially under noisy conditions and with an uneven station distribution. Operational monitoring of the highly active Santorini sequence faced all these challenges.

The onset and subsequent acceleration of reported low-magnitude seismicity that began on January 25, 2025, prompted the British Geological Survey (BGS) researchers of this group to initiate daily tracking using the QuakeFlow (QF; Zhu et al., 2023) DL-based workflow for earthquake catalogue development (Timeline *Team*, 31 January; Fig. 2) (refer to Supplementary Material, Machine Learning Event Detection). We note that the QuakeFlow workflow includes PhaseNet, a DL phase picker (Zhu and Beroza, 2019), and GaMMA (Zhu et al., 2022), an unsupervised earthquake associator based on a Bayesian Gaussian Mixture Model. Given the need to respond immediately, we deployed these existing methods. We chose to use the basic PhaseNet phase-picking model because we were familiar with it and it has been successfully deployed in the volcanic settings of Mayotte (Retailleau et al., 2022), Hawaii (Wilding et al., 2023), and Campi Flegrei (Tan et al., 2025) with good results that include the detection of hybrid and Long-Period (LP) events. In future deployments, other phase-picking models could be considered, such as Volpick (Zhong and Tan, 2024), which is designed to be effective for volcanic seismicity. GaMMA is an unsupervised approach for phase association so it should be effective provided the velocity model is sufficiently accurate.

On February 1, 2025, DL revealed previously unreported microseismicity dating back to November 2024 with 1,530 events versus 164 events in the standard catalogue (Fig. S2). From this point forward, we daily accessed the National Observatory of Athens (NOA) data node (Evangelidis et al., 2021) of the European Integrated Data Archive (EIDA) to acquire continuous waveforms from the 24 operating seismic stations within 100 km of Santorini. DL products, specifically P- and S- seismic phase picks generated by PhaseNet (Zhu and Beroza, 2019), were shared daily among researchers of this team via Amazon Web Services (AWS) through the Amazon S3 data exchange protocol to accelerate analysis including first order tasks, such as seismicity rate tracking using the daily DL catalogue, quality-check of results and event location accuracy (Timeline *Team*, 2 February; Fig. 2). This near-real time analysis was subsequently aided by additional seismic analysis such as moment tensor solutions, enhancement of the daily DL catalogue, searching for hybrid and LP events, and seismic tomography, to help identify the source of the unrest.

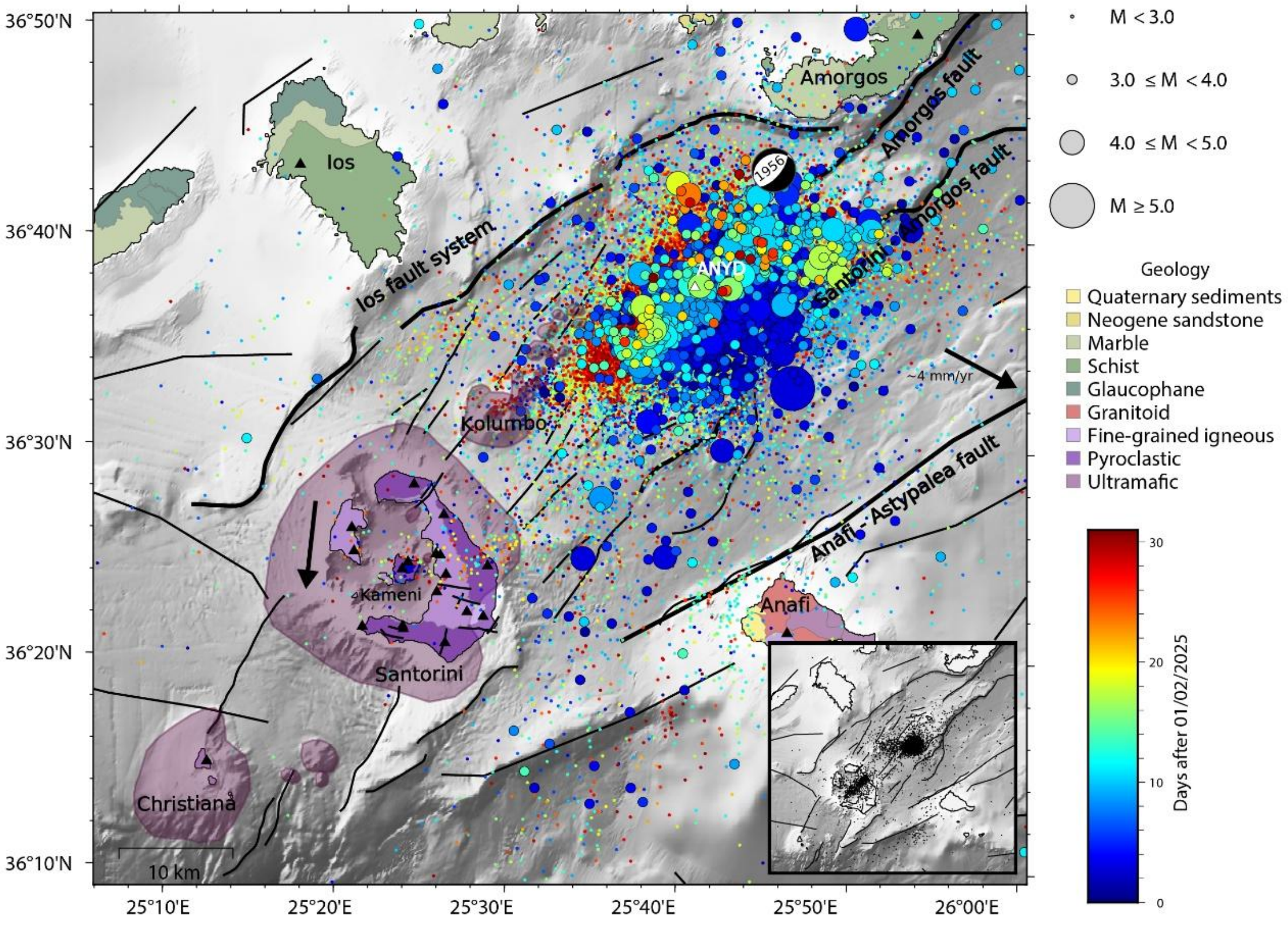

Figure 1: Overview of the 2025 sequence. Santorini volcano and the submarine Kolumbo volcanic chain, active faults and high-resolution bathymetry, as taken from the National Observatory of Athens (Ganas, 2025) and the European Marine Observation and Data Network (EMODnet Bathymetry Consortium, 2024), respectively. Black dots in the inset denote the seismicity observed between June 1, 2024 – January 31, 2025, whereas events that occurred from February 1 until March 3, 2025, are colour-coded by time. Seismicity within the aforementioned periods results from the implementation of the QuakeFlow (Zhu et al., 2023) DL workflow of this study. Geodetic vectors show the extensional strain field (McClusky et al., 2000). Seismic stations operating by the end of the seismic crisis are shown with black triangles. The black and white focal mechanism is that of the 1956 Amorgos mainshock (Brüstle et al., 2014). Dates are shown in DD/MM format. We also present in Fig. S1, the deviatoric moment tensors of events from this study with magnitude M > 4; using the same colour-map by time since February 1, 2025, mainly indicated normal faulting kinematics consistent with the main tectonic fabric of the region, but diverse combinations of reverse and strike-slip ruptures are also present.

Changes in seismicity rates are often the first and clearest line of evidence to signal the onset of magmatic unrest. In Santorini, the NOA seismic processing detected 3,733 events between $M_L$ = 0.5 and $M_L$ = 5.3 from February 1to March 3, 2025. Our DL approach detected 40,224 events during the same period.

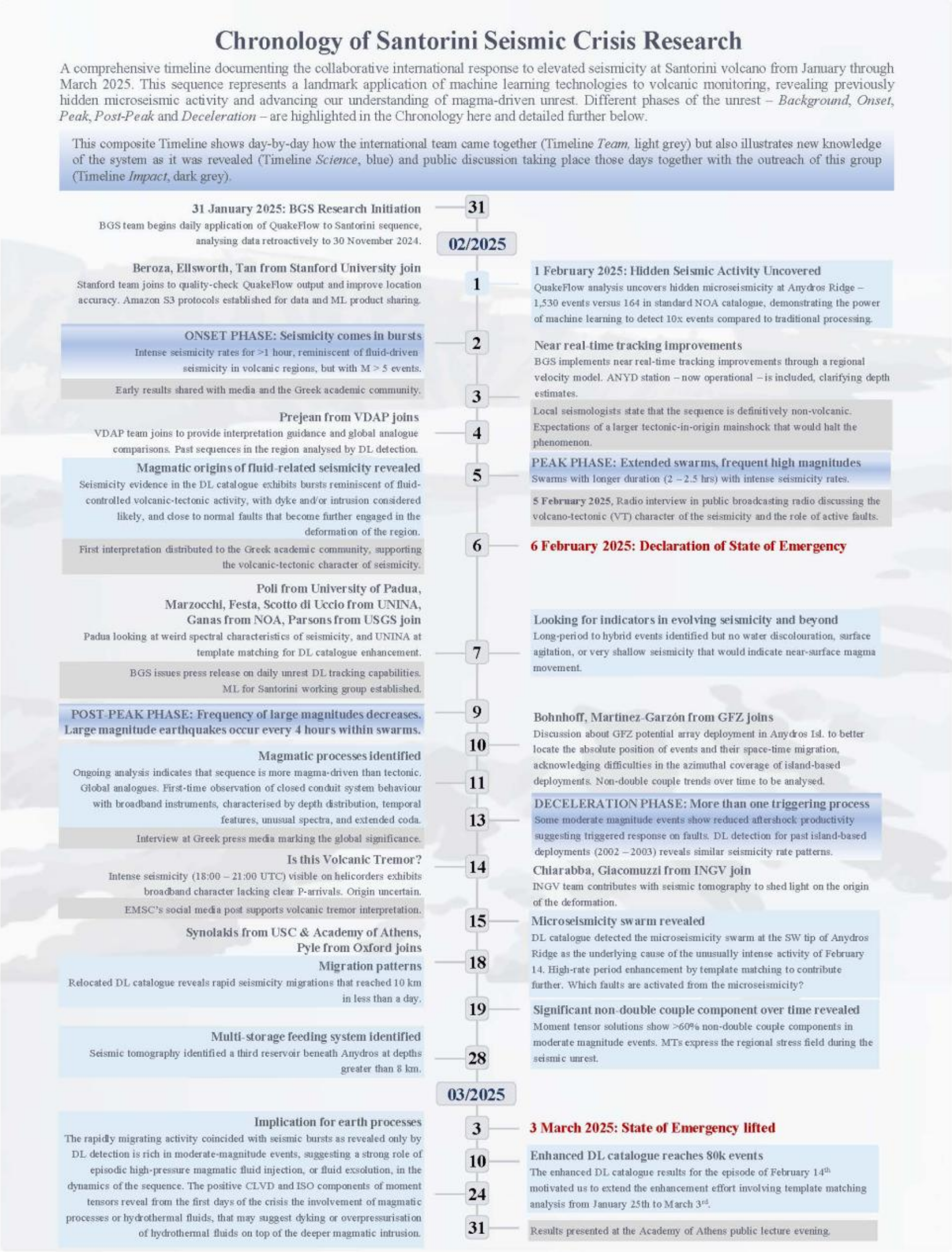

**Figure 2: Chronology of Santorini seismic crisis research (Part 1).**

**Figure 3: Chronology of Santorini seismic crisis research (Part 2).**

## February 2: A Signature of Fluid-Driven Volcanic Seismicity

The hourly earthquake count highlights the sudden increase of activity that lasted 14 days in early February (Fig. 4B). In the interval January 25–31, 2025, our DL catalogue revealed average hourly rates of only a few events (<15) per hour, but the onset of the widely felt seismic crisis on February 1st revealed a maximum of 68 events/hour. The average hourly number reported by the standard catalogue remained low (~6 events/hour above $M_L$=2.7; refer to Supplementary Material, Analysis of the Magnitude – Frequency Distribution), but DL detection (~60 events/hour above $M_L$=1.5) enabled the identification of emerging seismicity patterns, such as burst-like swarms (Hill et al., 1990) as early as 2 February 2, 2025 (Fig. 2). This was an important finding because spasmodic bursts of brittle failure earthquakes have long been observed in seismicity associated with volcanic and hydrothermal processes (Hill et al., 2002) but are not characteristic of tectonic processes. Individual spasmodic bursts have been interpreted as episodes in which fluids, exsolved from rising magma, move quickly through the Earth's crust, and trigger earthquakes (Shelly et al., 2015). In real-time continuous waveforms, those burst-like swarms were identified in helicorders (Fig. 3) ANYD station, February 4, 2025).

Repeated episodic bursts of seismicity have been observed at many volcanoes. When unaccompanied by an eruption, they are often interpreted as indicating instability in the shallow hydrothermal system (Fujita, 2008). Prior cases resemble the Santorini crisis seismicity in terms of their episodic nature but also differ in some ways. The spasmodic bursts at Santorini are unusual in terms of the large magnitude of earthquakes involved. The latter prompted us to look at the recent past of Santorini offshore region using DL to understand the time history of these crises (Timeline *Science*, 4 February; Fig. 2) and to compare those with global analogues aided by the experience of the Volcano Disaster Assistance Program (Timeline *Team*, 4 February; Fig. 2). During the ongoing crisis, we also processed 10 temporary stations deployed during the 2002–2003 CYCNET experiment (available through GFZ EIDA open data node in S3 protocol upon request to co-author Marco Bohnhoff who led this deployment) (Fig. S3 in Supplementary Material). We find 3,109 (9/2002 – 10/2003) against 127 events in the standard catalogue (9/2002 – 7/2004), as reported in (Bohnhoff et al., 2006). The DL enhancement of the past crisis, from 50 events per month to >50 events per day during the peak of the crisis, reveals previously unseen similar-to-2025 burst-like pulses near Anydros (green box; Fig. S3 in Supplementary Material) around March 2003; however, maximum magnitudes were considerably smaller than in 2025, reaching $M_L$ = 3.8.

The seismic crisis was characterised by an increased number of observed events within the first 14 days, expressed in Figure 4A and B. Overall, high seismicity rates during these burst-like swarms characterised the crisis, with median hourly rates following the occurrence of M4+ earthquakes of 65 (February 2–5, 2025), 122 (February 6–9, 2025) and 73 (February 10–13, 2025) events, reaching only a few events per hour after February 23, 2025. The maximum hourly rates per day are reported in Fig. 4B, reaching as high as 150 earthquakes during the peak of the crisis. In Figure 4A, the dashed black line represents the running mean of hourly rates, illustrating how the burst-like swarms were identified in the DL catalogue. These bursts of higher seismicity rates include the largest magnitude events of the seismic crisis. Many bursts began with a moderate magnitude event (M ~ 4), then escalated in magnitude within a few minutes to hours before subsiding. The duration of those burst-like swarms, as described by the inter-event times between successive moderate and larger events with $M_L > 4.0$, tended to increase over time (Timeline *Science*, 4 February; Fig. 2). When the seismic crisis intensified on February 3–4, 2025, the timing of larger magnitude events indicated that the duration of spasmodic bursts was just over an hour (76 min ± 23; 84 min ± 15), followed by an almost two-fold increase in swarm duration on February 5–6, 2025 (135 min ± 17; 105 min ± 17) (Timeline *Science*, 5 February; Fig. 2).

## February 6: Declaration of State of Emergency

The February 6, 2025, declaration of a state of emergency in Santorini demonstrated uncertainty over the evolving phenomenon and the hardships of the local population who were feeling almost continuous ground shaking (Timeline, 6 February; Fig. 2). In evaluating seismicity near volcanoes, it is critical to consider the character of the seismicity in addition to developing a detailed catalogue. The transition from purely brittle failure earthquakes to seismic events indicating the flux of fluids directly, such as LP earthquakes and tremor, is a key transition point used in eruption forecasting (Chouet, 1996; White and McCausland, 2019). For this reason, starting on February 7, 2025, during the *Peak* phase, we closely examined seismic data using both time series plots, spectrograms and frequency index (FI) estimations to guide our search for long-period earthquakes, volcanic tremor, or earthquakes with unusual source processes, and/or shallow seismicity that would indicate near-surface magma movement (Timeline *Science*, 7 February; Fig. 2). Our initial search for non-tectonic events failed to find definitive evidence of either LP earthquakes or harmonic tremor. Indeed, our preliminary calculations of the ratio of spectral energy from 1 to 5 Hz to the energy from 5 to 15Hz, which revealed a FI above 10 despite not being large magnitude (M < 3), supported the hybrid nature of seismicity (Greenfield et al., 2019). Despite these initial results, the temporal and spatial evolution of the crisis was unlike any historical analogue for a purely tectonic event. We further initiated offline catalogue enhancement with experts from the Volcano Disaster Assistance Program (VDAP) and the University of Napoli (Timeline *Team*, 7 February; Fig. 2), focusing on expert advice on the significance of LP events and advancing detection using waveform templates respectively.

## February 9: Moment Tensor Analysis

After February 8, 2025, the large magnitude events within the increased seismicity bursts began to spread out in time with larger inter-event times (~6.5 hours). From February 9, 2025 (Fig. 2) until February 12, 2025, spasmodic burst durations remained approximately 220 min, followed by a pronounced Omori law-type decay more evident after February 12, 2025, indicating a spatially broader pattern of seismicity triggered by the volcanic activity, as observed in other volcanoes globally (Florez et al., 2025). We could not exclude the possibility that seismicity triggering as a response to the local stress perturbation started earlier in the crisis before February 10, 2025, but that signature would have been masked by the much more intense volcanically driven seismicity. As moment tensors became available through the efforts of GFZ scientists beginning on February 10, 2025 (Timeline *Team*, 10 February; Fig. 2), we considered the time evolution of the M > 4 events of the sequence that continued to occur in the *Post-peak* phase to help address the role of magmatic versus tectonic processes during this sequence.

Deviatoric and double-couple moment tensors of events with magnitude M > 4 mainly indicate normal and oblique faulting kinematics consistent with the main tectonic fabric of the region. Typically, one of the fault planes strikes ~N45 – 60°E, in agreement with the main orientation of mapped faults in the region (Fig. S1). The distribution of P and T axes from the deviatoric solutions shows a predominance of near-vertical P axis orientations (Fig. S9a). When considering the full moment tensor solution, the proportion of strike-slip and reverse faulting increases slightly, resulting in several moment tensors with P axes close to horizontal (Fig. S9a). In the next step, we estimated the stress field from the deviatoric and full moment tensor solutions for the stress field orientation (Martínez-Garzón et al., 2014, 2016). In both cases, the orientation of the three principal stress axes was consistent within the uncertainties, with the trend and plunge angles being $S_1$ $[95,71]$, $S_2$ $[-122,15]$ and $S_3$ $[-29,11]$, representing an extensional normal faulting stress regime (Fig. S9b). This stress field orientation is consistent with the distribution of P and T axes of earthquakes in the Amorgos and Kolumbo volcanic chain (Friederich et al., 2014). Most of moment tensors for the larger events shown in Fig. S1 are associated with normal faults (e.g., a, b, e in Fig. S1), although some strike-slip (e.g. I, l, n, E, G in Fig. S1) and reverse mechanisms (e.g., d in Fig. S1) are also present. Reverse faulting as indicated by moment tensors was most notable early in the seismic crisis.

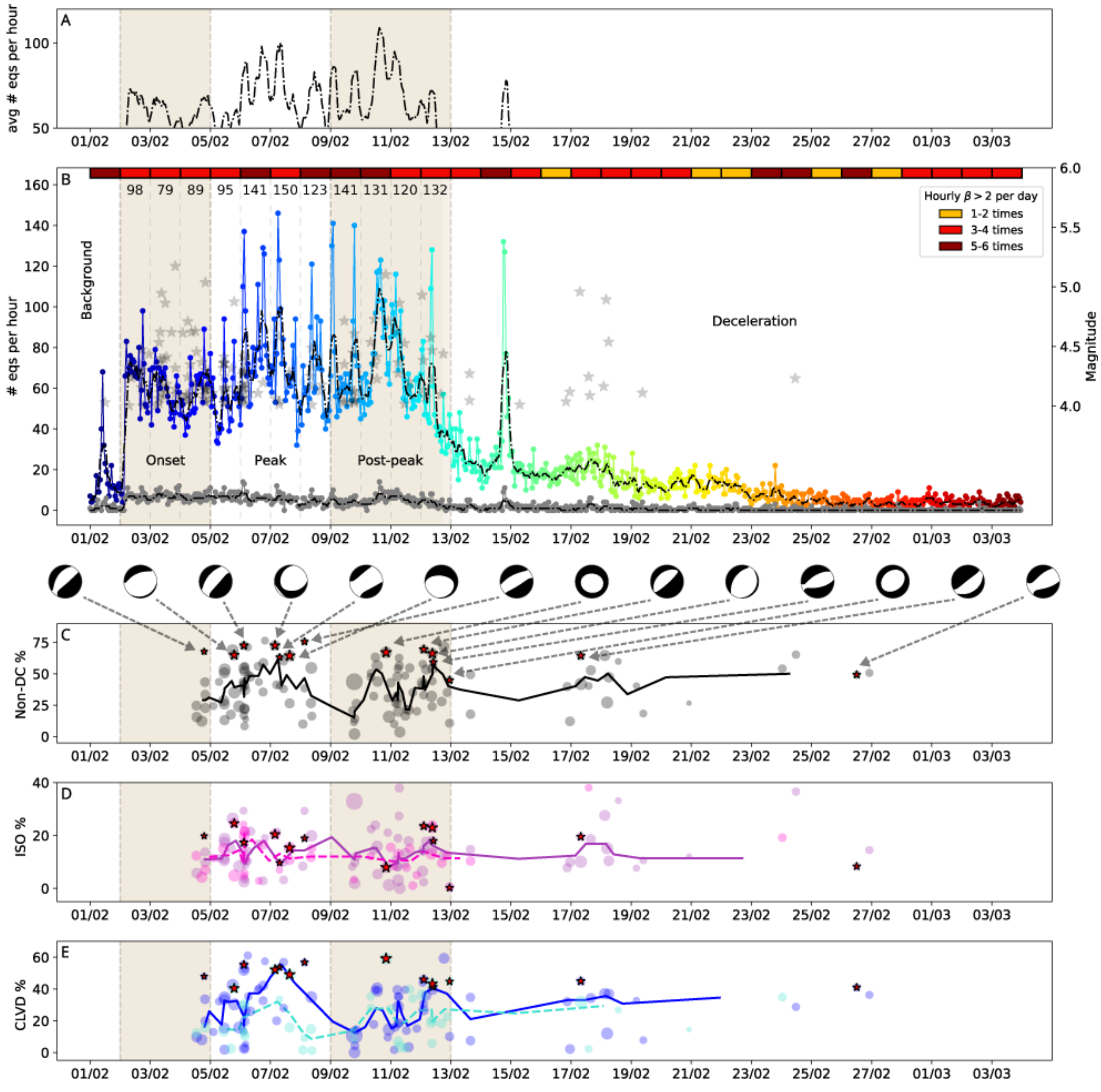

**Figure 4: Tracking seismicity phases during the crisis using machine learning for earthquake catalogue development. Vertical lines correspond to different phases based on seismic evidence. A) Average hourly count of DL-detected earthquakes. The black line denotes the median value with windows of 6 events and moving in steps of 1 event. B) Hourly count of earthquakes above the most frequently observed magnitude from DL detection (Zhu et al., 2023) (in colour) versus standard catalogue from National Observatory of Athens (NOA) (in dark grey), accessed daily. Stars**

**denote the time of occurrence of earthquakes with magnitude M ≥ 4.0. The maximum hourly rates per day are reported for the *Onset*, *Peak* and *Post-peak* phases. Using the DL-based catalogue, we calculate the $\beta$ statistic (Matthews and Reasenberg, 1988; refer to Supplementary Material, Seismicity Rate Change Quantification in hourly intervals to follow our seismicity rate tracking. We find 115 instances of $\beta$ > 2, indicating significant hourly seismicity rate increases (Hill and Prejean, 2015; Marsan and Wyss, 2011). The daily number of instances with significant hourly rate increase is shown at the top of the figure. During the *Peak* and *Post-Peak* periods, we had at least two days with more than 5 instances of significant hourly rate increases. C) Non-double couple (non-DC) proportion of individual moment tensors (circles size-encoded with their magnitude). Red stars denote events with absolute CLVD > 40% and plunge of P-axis > 60°. The full moment tensors of those events are displayed on top. D, E) Same as C), but for the absolute values of the isotropic (ISO, D) and compensated linear vector dipole (CLVD, E) components. In D), purple and pink circle and line colours refer to positive and negative ISO values respectively. Similarly, in E), blue and cyan circles and lines refer to positive and negative CLVD values respectively.**

Non-DC mechanisms including large compensated linear vector dipole (CLVD) components have been found in various environments hosting volcanic processes such as fluids propagating through dyke intrusions, explosive or implosive sources, demonstrated in isotropic (ISO) components. Notable examples include the Long Valley Caldera, California (Dreger et al., 2000; Julian, 1983), the Miyakejima Island, Japan (Shimizu et al., 1987), the Eyjafjallajökull volcano and the Hengill – Grensdalur volcanic complex in Iceland (Dahm and Brandsdóttir, 1997; Miller et al., 1998), Campi Flegrei in Italy (Campus et al., 1996; Campus and Cespuglio, 1994), and preceding the 1991–1993 Etna eruption (Saraò et al., 2001). In volcanic regions, these significant non-double couple (DC) components in moment tensors are usually attributed to cracks opening and cavities collapsing in a heterogeneous stress field after a major dyke injection episode. Global studies have shown that the presence of large non-DC components in moment tensors of earthquakes near volcanoes are frequently associated with eruptive processes (Pesicek et al., 2021; Shuler et al., 2013). Non-DC components are also present in geothermal environments, such as in the Larderello geothermal area, Italy (Kravanja et al., 2000) or The Geysers geothermal field in California (Martínez-Garzón et al., 2017). Here, we estimated both the deviatoric (trace-null) and full moment tensors by means of an approach using the sign and amplitude of the P-wave first motion (Kwiatek et al., 2016; refer to Supplementary Material, Moment Tensor Calculation).

Past studies suggest that volcanic and geothermal earthquakes (Julian, 1983; Panza et al., 1993) may involve non-DC mechanisms. Depending on the moment tensor (MT) decomposition, the non-DC components may be a combination of CLVD and ISO components (full moment tensor), or they both may be represented by the CLVD component (deviatoric moment tensor) (Knopoff and Randall, 1970). We used both full moment tensors and deviatoric moment tensor solutions to identify temporal variations of the non-DC component, which exceeded 40% during two episodes February 5–8 and 10–12, 2025). During February 5–7, 2025, we observed a progressive increase in the non-DC component from about 20% to almost 60% (Fig. 4C). This increase came mainly from the CLVD component (Fig. 4E), while the ISO component (Fig. 4D) only increased by approximately 10%. During February 10–12, 2025, we observed a similar evolution, with a progressive CLVD increase and minor variations in the ISO components, representing both explosions and implosions. In these periods, we observed moment tensors that reached 60% non-DC, predominantly exhibiting normal slip mechanisms with CLVD > 40% and P-axis plunge > 60° (denoted with red stars in Fig. 4C), D) and E). Such moment tensors with remarkable CLVD components and almost vertical P axes have previously been identified in volcanic dyke intrusions (Shuler et al., 2013).

The evidence from seismicity rates and non-DC components of moment tensors indicated that magmatic processes were driving evolving seismicity (Timeline *Science*, 11 February; Fig. 2). In Figure 3 (Chronology), we describe in detail the distinct *Background*, *Onset*, *Peak*, *Post-peak* and *Deceleration* phases, identified by all data as they became available.

## February 11: Search for LP Events, Shallow Seismicity and Unusual Spectra

During the crisis, we searched for LP-enriched earthquakes using two methods. First, we visually examined the continuous seismic data on station ANYD and identified seismic events that seemed to be enriched in low frequencies (Fig. 5). We identified 22 events that were enriched in LP energy and lacked high-frequency energy. We used these events as templates to search for additional similar events recorded at ANYD using a cross-correlation threshold of 0.7. We identified 39 additional LP events for a total of 61 during the month of February. Five of the identified LP events were included in the NOA catalogue. These events were located between 5 and 12 km depth on the NW edge of the seismicity band at a distance of roughly 15 km from Kolumbo. Temporally, these events occurred throughout February at a maximum rate of 7 detectable events per day. We did not correct for path effects but noted that other nearby earthquakes of similar size had more high-frequency content in the waveforms recorded at ANYD. Thus, it is likely that the frequency content indicates source processes rather than path effects. Second, we searched for LP-enriched earthquakes in a more comprehensive manner using the Frequency Index (FI) (Buurman and West, 2010). We computed the FI for all events in the publicly available catalogue published by NOA, using a low frequency band of 1–2.5 Hz and a high frequency band of 5–10 Hz (Fig. 5) to search for outliers. This analysis revealed a broad range of FI (Fig. 5). Because seismicity occurred over a wide area and this is a single station analysis that does not incorporate path effects, additional analysis would be required to interpret results with confidence. However, to first order, some events seemed to have had unusually low frequency content. Papadimitriou et al. (2015) observed LP events under Santorini during the 2011–2012 unrest and interpreted them as evidence of fluid migration.

The LP analysis conducted was not a comprehensive study of potential LP-enriched earthquakes. Rather, the LP analysis was a survey of how prevalent these events might be to assess the character of the intense seismicity unrest as it was occurring. Relatively few potentially LP earthquakes were found, and we discerned no shallow volcanic tremor due to near surface resonance of fluid filled cracks. This supported our inference during the crisis that eruption was not imminent. However, we interpreted the occurrence of LP earthquakes in combination with the unusual time series of the brittle-failure seismicity as supporting the hypothesis that the swarm was driven by high-pressure fluids from a magmatic intrusion.

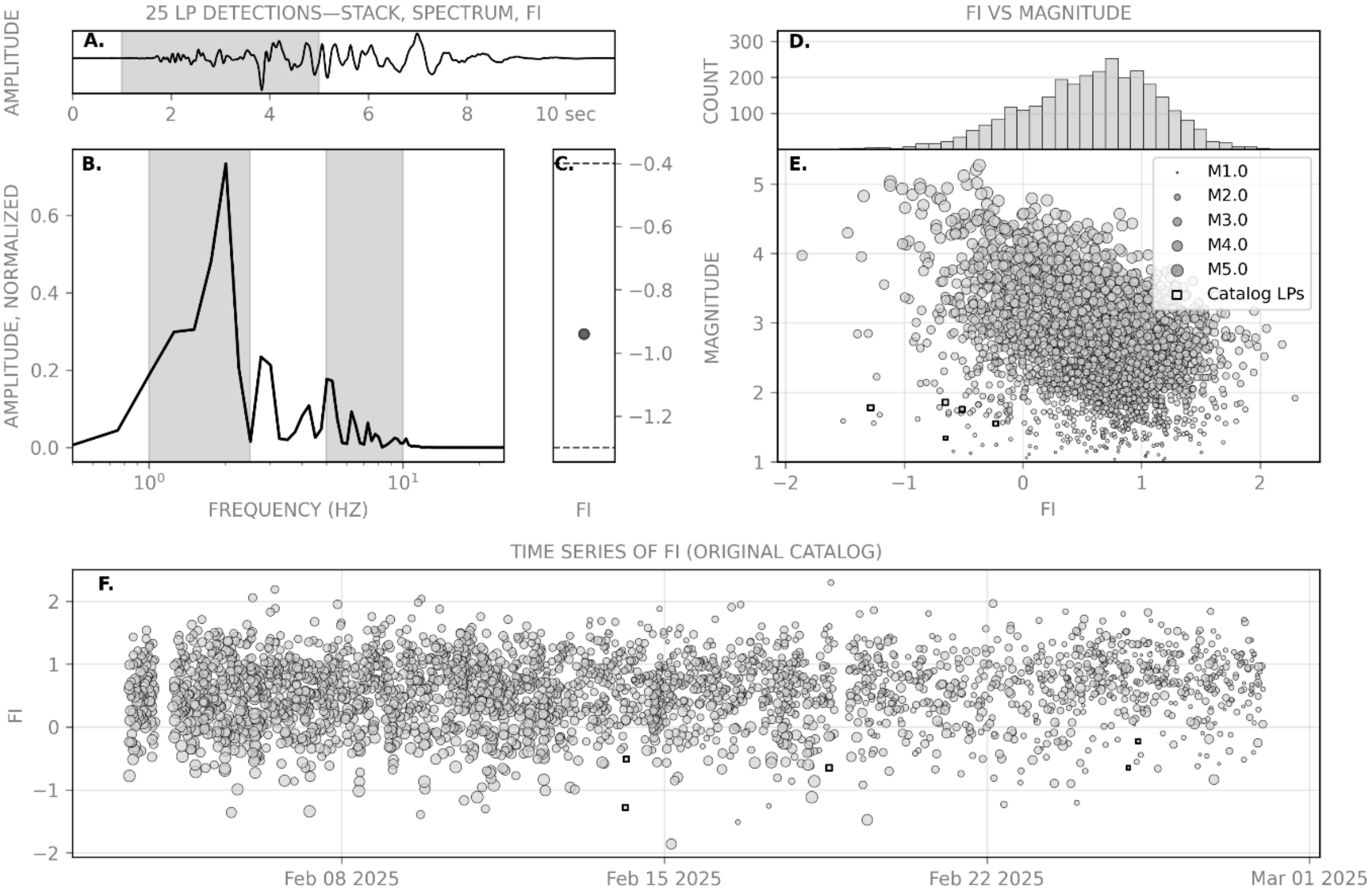

**Figure 5: Long-Period earthquakes (LPs) within the unrest. A) Stacked waveform of 25 LPs with an average cross correlation coefficient of 0.7. B) Spectra of stack with frequency bands used to compute the Frequency Index shown (1 – 2.5 Hz and 5 – 10 Hz). C) Frequency Index of stack. This event had a low Frequency Index (FI), indicating that it is enriched in long-period energy. D) Histogram of FI. E) Relationship between FI and magnitude using the standard**

**catalogue. Black squares show LPs we detected, which are included in the catalogue. F) Time series of FI for catalogue events.**

## February 11: A Search for Global Analogues

In addition to the uniqueness of the seismicity bursts, the sheer number of M 4 – 5 earthquakes, which continued for weeks without a typical Omori decay, was remarkable. Earthquake sequences with characteristics similar to this are very uncommon in the history of seismic monitoring (Belachew et al., 2013; Cesca et al., 2020; Gudmundsson et al., 2014; Neal et al., 2019; Uhira et al., 2005). Figure 6 shows the global analogues we considered in February 2025 during the crisis (Timeline *Science*, 11 February; Fig. 2). In Figure 3 (Chronology), we highlight some aspects of volcanic seismicity observed elsewhere. All analogues either involve eruptions or magmatic intrusions and are specifically associated with basaltic dyking, though the speed of dyke migration and location of earthquakes with respect to the dyke vary between the analogues. We are not aware of any purely tectonic sequences that display these characteristics. For these reasons, we inferred in early February 2025 that the Santorini crisis likely involved magmatic dyking.

It is important to note that the analogues in Figure 6 either involve rift zones or locations with mafic volcanism due to upwellings of primitive magmas from depth. None of these modern analogues involved more felsic explosive volcanic systems, in which eruptions are triggered by deep mafic intrusions, as is thought to occur near Santorini. This observation likely results from the short history of modern seismological monitoring and highlights how some aspects of the Santorini crisis were unprecedented.

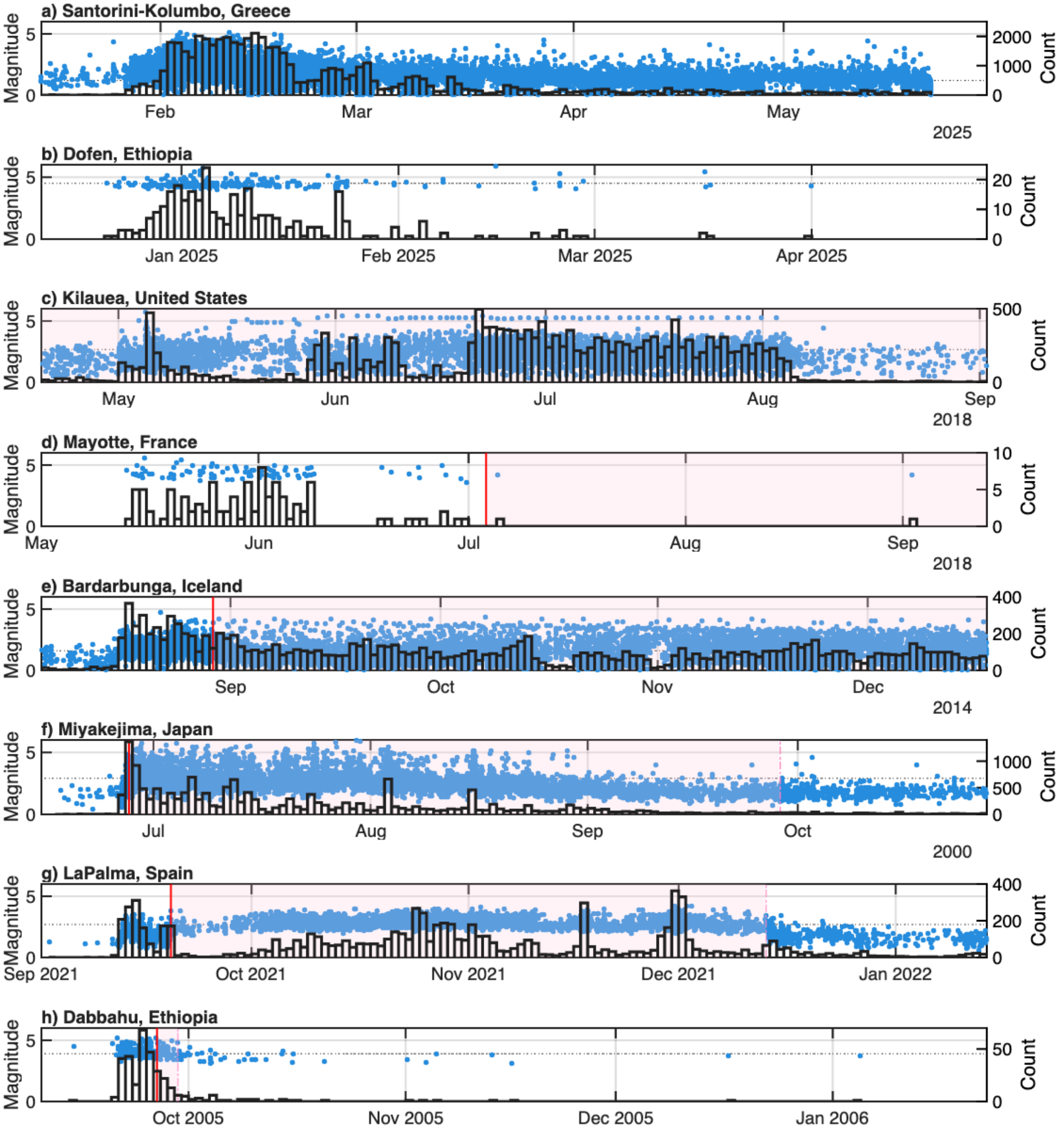

**Figure 6: Comparison of earthquake magnitudes and rates accompanying select dyke intrusions: a) this study, earthquakes are shown within r = 50 km of Kolumbo volcano; b) Dofen, Ethiopia, r = 100 km, data from USGS ComCat, c) Kilauea, Hawaii, r = 50 km, data from USGS ComCat; d) Mayotte, France, r = 60 km, data from ISC; e) Bardarbunga, Iceland, r = 50 km, data from SIL; f) Miyakejima, Japan, r = 50 km, data from JMA; g) La Palma, Spain, r = 50 km, data from IGN; h) Dabbahu, Ethiopia, r = 50 km, data from ISC. Radii (r) refer to distance from the locations given by the GVP. Eruptions are shown in red and sourced from the GVP. For b) Dofen, no eruption occurred but dyke intrusion is well documented (Lewi et al., 2025). Data sources are further described in (Pesicek et al., 2021) and the Data Availability statement.**

## February 13: Deceleration Phase and a Review of Recent Past Activity

On February 12, 2025, the first Global Navigation Satellite System (GNSS) station on Anydros Islet was installed to complement existing geodetic stations on Santorini. As the sequence entered the deceleration phase (Timeline *Science*, 13 February; Fig. 2) with evidence indicating fault triggering as a response to local stress perturbations, we turned to the results of the DL-based reprocessing of the 2002–2003 crisis to address the question of how that past crisis halted. We processed 10 temporary stations deployed during the 2002–2003 CYCNET deployment (available through GFZ S3 data exchange) and permanent stations within 100 km distance. The standard catalogue has 127 events (September 9, 2002–July 7, 2004), whereas this study finds 3,109 (September 9, 2002–October 10, 2003). The DL enhancement of the past crisis increased the number of detected events from 50 events per month to >50 events per day during the peak of the crisis, revealing previously unseen but similar-to-2025 (Timeline *Science*, 13 February; Fig. 2) burst-like pulses near Anydros (green box; Fig. S3) around March 2003. This earlier sequence, however, had considerably smaller maximum magnitudes than in 2025, reaching a maximum of only $M_L$ = 3.8.

## February 14: The Need for Enhancing the Deep Learning Catalogue

The occurrence of repeated episodic bursts of seismicity on February 14, 2025, presented an even more demanding event detection environment for DL algorithms for multiple reasons (Timeline *Team*, *Science*, 14 February; Fig. 2). First, like other algorithms, DL algorithms struggle to identify events when inter-event times are very short such that waveforms strongly overlap (Park et al., 2023; Pita-Sllim et al., 2023), yet accurately characterising this type of activity rapidly is critical for informing emergency response decision making. Second, even if picks are made properly, it is challenging to associate them, i.e. to connect wave arrivals unambiguously to the events that generated them. Third, all these challenges are magnified by the limited spatial distribution of data across nearby islands. The detection of low-magnitude events in such a situation can be greatly enhanced using similarity-based approaches (e.g. template matching; Chamberlain et al., 2017), with events detected in near real-time using QF (Fig. 1) as the template set (Scotto di Uccio et al., 2023). We call the dataset here the *enhanced* DL catalogue. The combination of shorter time windows in template matching and narrow-band filtering of seismic waveforms allows for increased resolution of high-seismicity intervals, especially when using enriched template sets (Scotto di Uccio et al., 2023; refer to Supplementary Material, Catalogue Enhancement: Template Matching).

To illustrate the different phases of the seismic crisis, we select four examples as representative cases marking the evolution of the seismic crisis. We report on 30-min seismograms and spectrograms from station ANYD (here, component HHZ) for February 3 (Fig. 7a), 6 (Fig. 7b), 14 (Fig. 7c) and 17, 2025 (Fig. 7d), some of which include relatively large magnitude events within bursts, within the chaotic microseismic swarm and neighbouring fault activity that more closely resembles tectonic seismicity respectively. For example, in a 30-minute segment of seismic records during the high-seismicity episode on February 14th (Fig. 7c), we observed an increased number of closely spaced earthquakes with inter-event times of just a few seconds. The DL catalogue revealed a microseismicity swarm near the southwestern tip of the Anydros horst, which corresponded to the episode between 18:00 and 21:00 UTC on February 14$^{th}$ (Timeline *Science*, 15 February; Fig. 2). For the 210-second window between 18:18:10 and 18:21:40, our DL catalogue reports 8 earthquakes, whereas the enhanced DL catalogue, using the DL events as templates, resolves 16 events with magnitudes ranging from $M_L$ 1.8 to 3.4. Within the swarm period on February 14th (18:00 – 21:00, Fig. 7c), the enhanced DL catalogue identified 837 earthquakes with a median inter-event time of 11s – nearly doubling the number detected by DL alone (423 earthquakes).

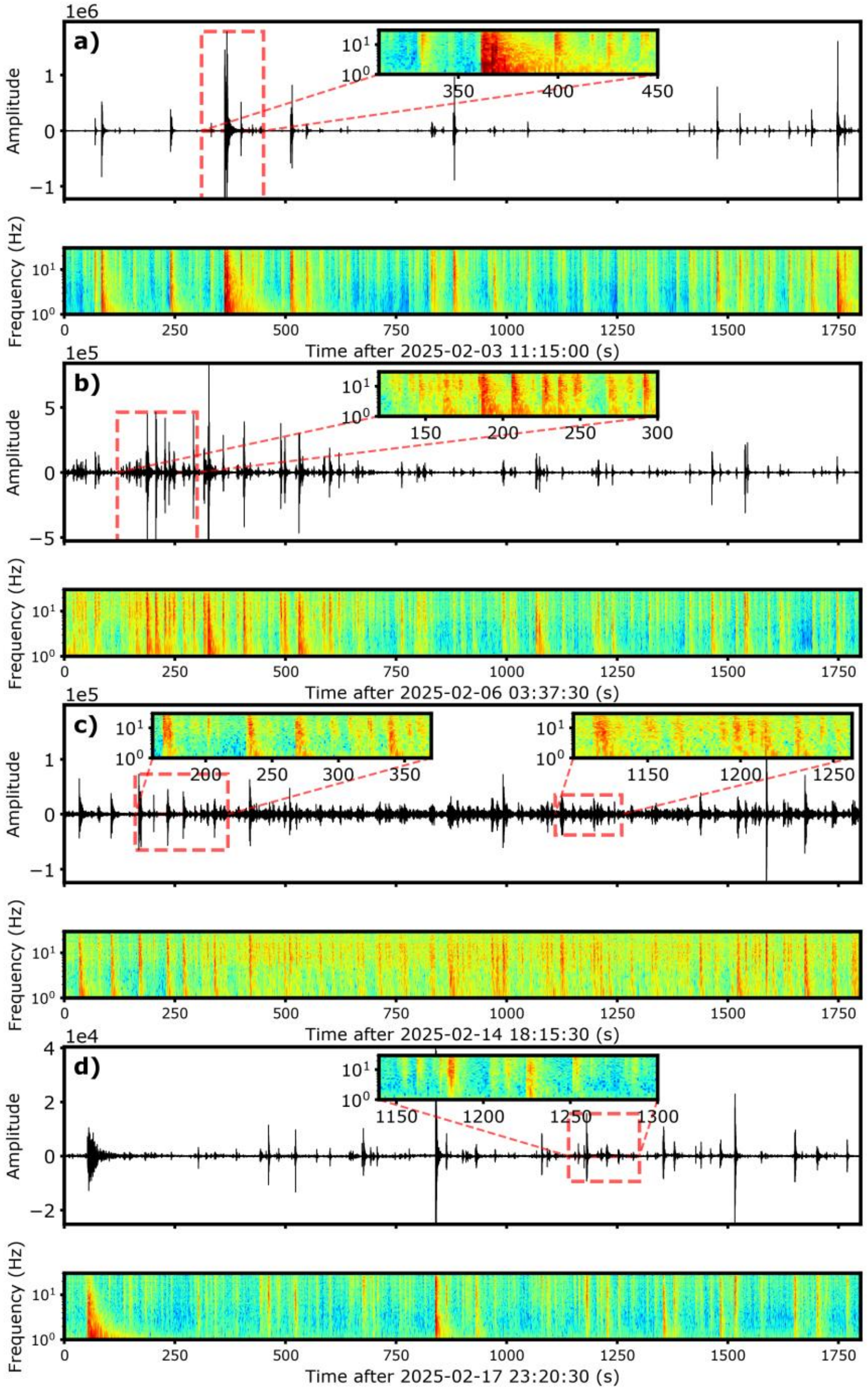

**Figure 7: Waveforms (Amplitude in counts) and spectrograms (Frequency in Hz) extracted for 30 minutes of waveforms at station ANYD (component HHZ) during high seismicity time windows that occurred on February 3 (panel a), February 6 (panel b), February 14 (panel c) and February 17 2025 (panel d). Vertical bands in the**

**spectrograms reveal the occurrence of nearly overlapping earthquakes, supporting the need for the integration of template matching detection techniques by leveraging the DL-based catalogue.**

The template-matching-enhanced DL catalogue consistently reports about twice as many earthquakes as the DL catalogue, primarily due to its higher detection resolution for events occurring very closely spaced in time. Figure S12 shows the temporal distribution of seismic events, along with the ratio of the number of earthquakes identified by template matching to those detected by machine learning in 1-hour bins. The time distribution highlights a consistent two-fold increase in the number of earthquakes introduced by template matching, with local maxima around the high-seismicity rate period, highlighted as red, described in Figure S12.

We consider the magnitude–frequency distributions (MFD) of the available catalogues. Herrmann and Marzocchi (2020) revealed that there are significant departures from the exponential-like magnitude–frequency distribution (MFD) for high-resolution catalogues in smaller magnitude ranges, suggesting that using a traditional magnitude of completeness may be misleading. Here, we prefer the term "most frequently observed magnitude" to describe the MFD. In the left panel of Figure S13, we compare the MFD of the enhanced DL catalogue (green) with the MFDs of our DL (blue) and NOA (yellow) catalogues. From these curves, we derived the most frequently observed magnitude, which relates to the slope change in the distribution (Wiemer and Wyss, 2000). The NOA catalogue reported a corresponding magnitude of $M_L$ 2.7. In contrast, the DL catalogue contains 10 times more low-magnitude events, reducing this value to $M_L$ 1.5 (Fig. S13). By combining deep learning with similarity-based approaches, the discussed magnitude was further reduced to $M_L$ 1.3 (Fig. S13).

Although the MFDs of the QF and enhanced DL catalogues (blue and green dots in the left panel of Fig. S13, respectively) seem similar, plotting the discrete number of earthquakes as a function of magnitude on a linear scale revealed different populations in the lower magnitude bins (Fig. S13, right panel).

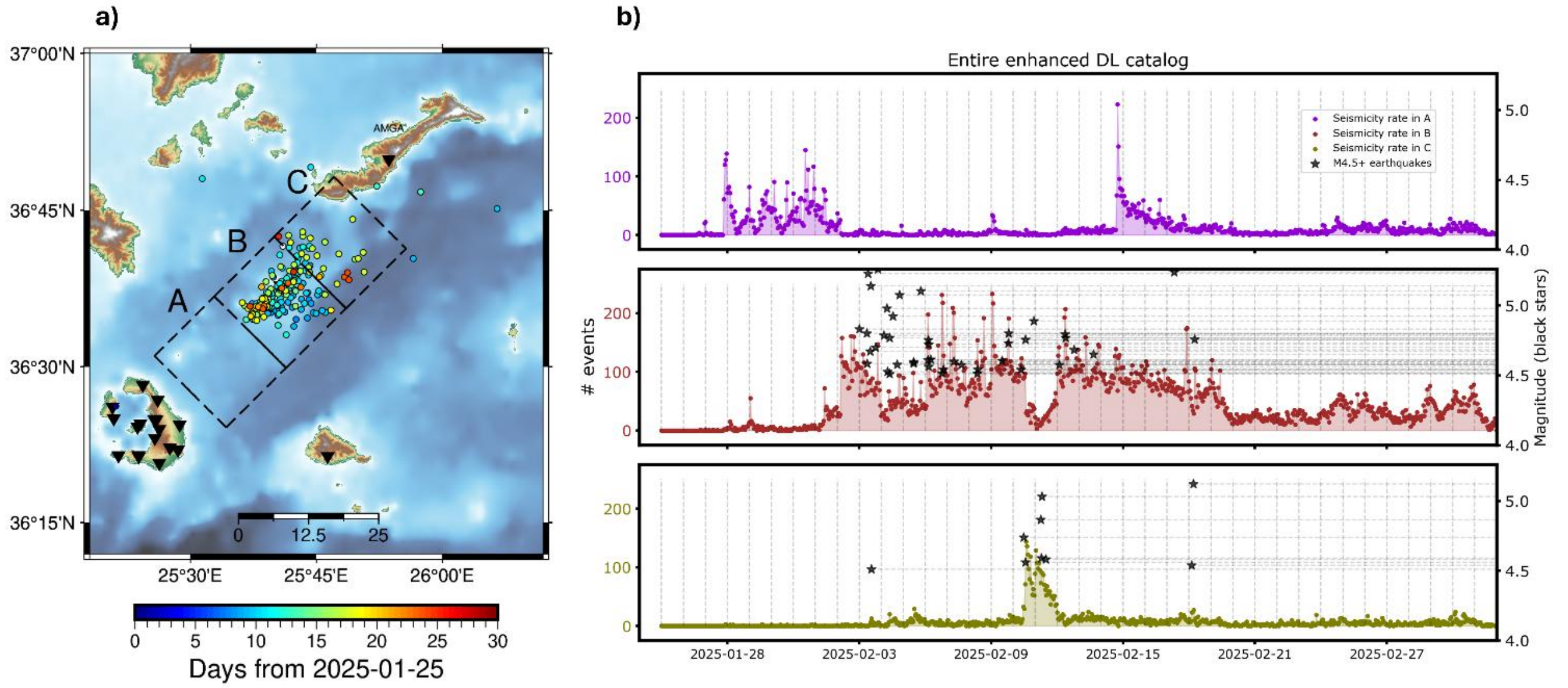

**Figure 8: Migration of microseismicity revealed by the enhanced DL catalogue. Left panel: spatiotemporal evolution of the M > 4.0 earthquakes. Dashed boxes, each spanning 15 km along the strike direction, mark the investigated areas at the southwestern tip of the Anydros ridge (Sector A), Anydros (Sector B) and Amorgos (Sector C). Right panel: earthquake count per hour in the individual A, B and C sectors (top to bottom). We highlight M > 4.5 earthquakes with stars, whose magnitude range is represented on the right vertical axis.**

Enhancement of the DL catalogue clarified the migration pattern of microseismicity involved in the evolution of the crisis (Fig. 8) (Timeline *Science*, 15 February; Fig. 2). During the Onset period, the seismicity rate in sector A (Fig. 8) strongly decreased and became more intense in sector B, bounding Anydros, with the occurrence of M > 5 earthquakes. Seismicity migrated farther NE, as indicated by the increase in the number of earthquakes in sector C and supported by higher S – P travel times observed at Santorini stations. In the *Peak* phase, the high seismicity rate beneath Anydros (sector B) is characterised by the occurrence of >200 earthquakes per hour. *Post-peak* epicentres suddenly migrated 15 km farther NE from the latest positions on February 9th, with M > 5 earthquakes and high productivity in sector C. In the *Deceleration* phase, seismicity became strongly productive near Anydros, including large magnitude earthquakes up to February 17, 2025, when the seismicity decreased to low rates even after the occurrence of M > 4 earthquakes. We expect that tectonically loaded normal faults would have lower stressing rate than volcanic bursts, which is expected to manifest in triggered seismicity quite differently (Dieterich, 1994). Seismicity rates in the *Deceleration* phase (post-February 12$^{th}$) are possibly related to the stress response of tectonic faults in the area (Timeline *Science*, 15 February; Fig. 2).

The question that emerged at that point was: which faults are (re)activated and what are their characteristics?

## February 15: Fault Activity

We relocated all earthquakes associated by GaMMA (Zhu et al., 2022) from November 30, 2024to March 10, 2025 using TomoDD (Zhang and Thurber, 2003), incorporating a 1D velocity model inverted with VELEST (Kissling et al., 1994, 1995), resulting in 57,886 relocated events (Timeline *Science*, 15 February; Fig. 2). From this, we selected the 36,280 best-located earthquakes and all events with a magnitude larger than 3.

The relocated seismicity revealed two subparallel, northeast to southwest (NE–SW)-trending alignments of seismicity within the Anydros horst (Fig. 9A). The southeastern strand passes through Anydros and links the Anydros fault system in the southwest with the Amorgos fault system in the northeast. This structure exhibits a predominantly northwest-dipping geometry, with dips decreasing from near-vertical in the central section to shallower toward the northeast (Figs. 9B, 9C–H). Hypocentres are concentrated from 5–10 km depth, deepening slightly towards the southwest. Southwest of Anydros, a cluster of events reached near-surface depths (Fig. 9E), whereas a separate cluster north-northeast of Kolumbo volcano extended to ~15 km (Figs. 9B, 9C). The northwestern strand displays a more variable geometry. In its southwestern part, seismicity indicates a relatively steeply dipping structure, whereas in the central and northeastern sections, near-horizontal, southeast-dipping features are visible (Figs. 9E–F). Seismicity here is shallower and restricted, mostly at 5–9 km depth. Noticeably, both strands show significant along-strike variability, ranging from narrow, well-defined planes to diffuse zones likely representing intersecting faults or distributed damage. In addition to the two sub-parallel NE–SW-trending systems, we also observed some activity on the Santorini–Anafi fault system (Fig. 9A). This structure produced numerous moderate-magnitude events, although its activity remained minimal compared to the Santorini–Amorgos system.

The relocated seismicity clarified the temporal evolution of the seismic unrest that showed pronounced and punctuated along-strike migration (Fig. 10) (Timeline *Science*, 19 February; Fig. 2). The seismic unrest initiated on January 28, 2025 (Phase 0), ~8 km northwest of Kolumbo volcano at 10–15 km depth (Figs. 10A–B). Between January 28 and February 1, 2025 (Phase 0), activity migrated locally to shallower depths and south-eastward, where a small swarm was observed on February 1–2, 2025 (*Onset*). Within the *Onset* and early *Peak* phase (February 2–6, 2025), seismicity rapidly propagated ~20 km northeast in a series of bursts along the Anydros ridge, passing beneath Anydros and reaching the southwestern tip of the Amorgos fault system (Figs. 10D–E). Within the late *Peak* phase (after February 6, 2025), activity shifted back to the southwest, but with more earthquakes concentrated at depths >10 km. In the *Post-peak* phase (February 10, 2025), another major north-eastward migration-initiated northwest of Anydros at ~10 km depth, propagating another ~10 km northeast in less than a day, then back-propagating toward previously activated zones (Figs. 10F–G). The final prominent migration occurred on February 14, 2025, characterised by an upward- and northeast-migrating microseismic swarm at the southwestern tip of the activated structure (Figs. 10A, 9H–I).

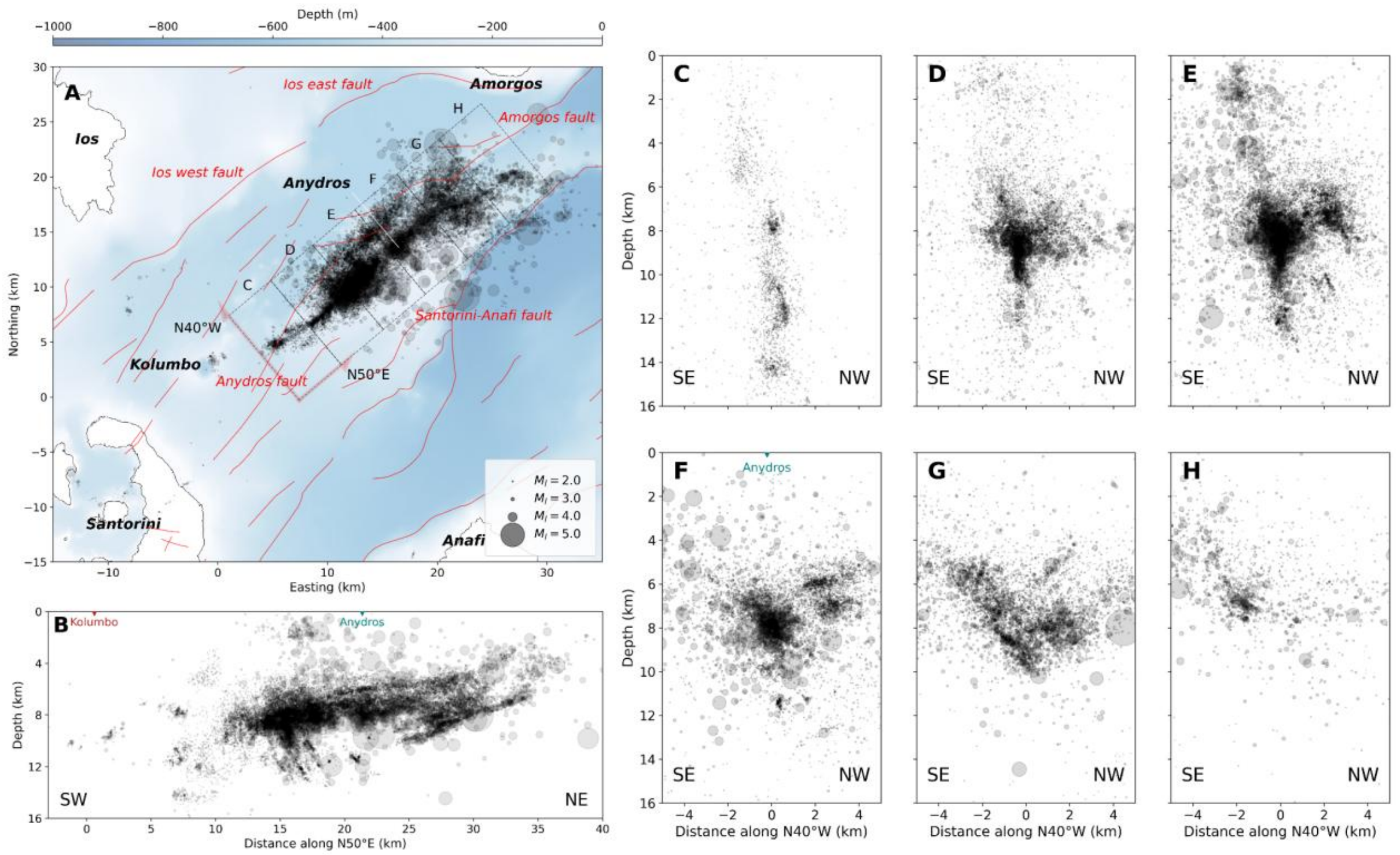

**Figure 9: Earthquake spatial distribution after TomoDD relocation, November 30, 2024, through March 10, 2025. A) Map view of relocated seismicity (black circles) as we determined it at the time and mapped faults (red lines). Circle sizes scale with magnitude. B) Cross-section along N50°E illustrating the overall depth distribution. C–H) Sequential cross-sections along N40°W, corresponding to the locations labelled in panel A. Only 36,280 best-located earthquakes and all events with a magnitude larger than 3 are shown in this figure.**

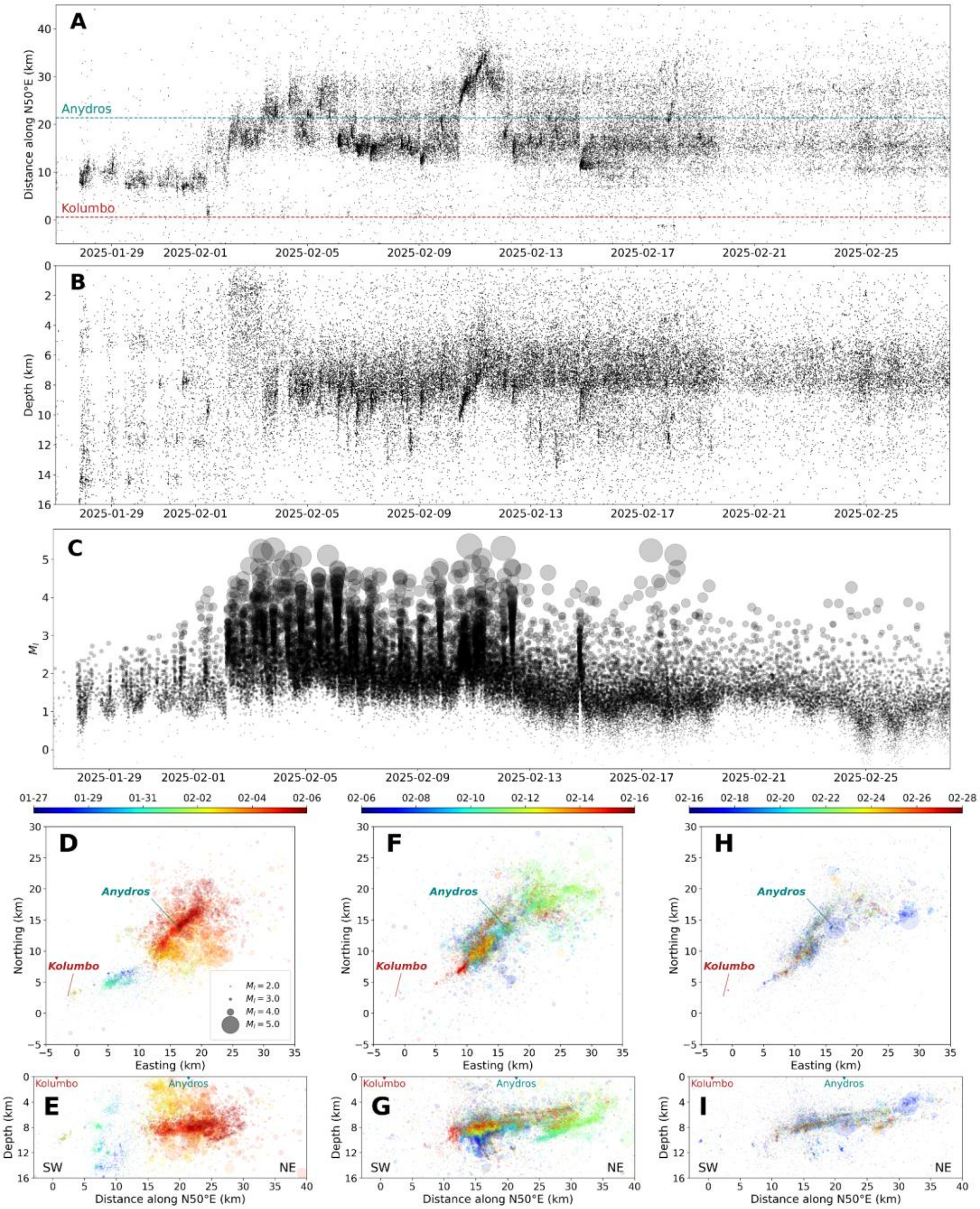

**Figure 10: Spatiotemporal evolution of seismicity from January 27 to February 28, 2025. A, B, C) Temporal progression of earthquake locations (projected along N50°E), depth, and local magnitude. D, F, H) Map views of seismicity for three successive time intervals: 27 January–6 February; 6–16 February; and 16–28 February. Circle sizes scale with magnitude, and colours indicate origin time. E, G, I) Corresponding cross-sections along N50°E for the same intervals shown in panels B, D, and F. Note that the colour scale is consistently from older (blue) to newer**

**(red), but differs for panels D–E versus F–G versus H–I. Only best-located earthquakes and all events with a magnitude larger than 3 are shown in this figure.**

At this point in our timeline, seismic evidence suggested that volcanic-tectonic activity resembling dyke formation characteristics occurs close to normal faults that become further engaged in the deformation of the region with seismicity spatially migrating along strike. Together with the significant non-DC components of events above M > 4, the above information indicates a strong role for episodic high-pressure magmatic fluid injection or fluids exsolution in the dynamics of the sequence (Timeline *Science*, 18 – 19 February; Fig. 2).

## February 28: Tomographic Imaging

Crustal structure discerned through tomographic imaging can inform models of the seismic crisis (Timeline *Science*, 28 February; Fig. 2). We used a non-linear Bayesian trans-dimensional inversion method (Giacomuzzi et al., 2024; Piana Agostinetti et al., 2015) to jointly estimate the P-wave velocity ($V_p$), $V_p/V_s$ ratio, and earthquake locations (refer to Supplementary Material, Seismic Tomography). The increased number of detected events and S-wave picks provided by DL enhances tomographic studies that are typically conducted either using primarily P-wave data (Autumn et al., 2025; McVey et al., 2019) or with a limited number of S-phase arrivals (Hufstetler et al., 2025). This enhancement enables the computation of velocity models with comparable resolution in both $V_p$ and $V_s$, and thus a robust estimation of the $V_p/V_s$ ratio that can help illuminate melt/fluid reservoirs. This novel tomographic approach is self-adapting and data driven. Model parametrisation, independence on starting reference velocity models, and the lack of global regularisation of the inverse problem, enhance the spatial resolution of the velocity models and reduce smearing of seismic anomalies. This approach is particularly effective at modelling seismic crises for scenarios with strongly uneven distributions of seismic events and stations, as in our case, where most stations are located on Santorini and most events occurred beneath the Kolumbo volcano and Anydros regions, resulting in raypaths with a dominant SW–NE direction.

In Figure 11, we present the mean velocity model derived from an ensemble of 5,000 model samples of posterior density distributions obtained by considering 13k well-located events of the DL catalogue (161,122 P-phases, 147,403 S-phases; refer to Supplementary Material, Table S2). We estimated picking uncertainties based on PhaseNet classification probabilities (refer to Supplementary Material, Seismic Tomography section). Map views of the velocity models and uncertainty estimates are provided in the Supplementary Material (Fig. S14–S16) and are below 0.4 km/s for $V_p$ and 0.2 for $V_p/V_s$ for most of the discussed seismic features.

Melt storage zones usually result in reduced velocity compared to the host rocks. Moreover, due to the stronger sensitivity of $V_p$ compared to $V_s$ to variations in fluid compressibility (Dvorkin et al., 1999; Takei, 2002), the $V_p/V_s$ ratio can be used to discriminate between volumes where magmatic fluids in a gas or supercritical fluid form (with very low $V_p$ and low $V_p/V_s$ ratio) or melt (with low $V_p$ and high $V_p/V_s$ ratio) prevail. The tomographic results reveal three main and distinct low-$V_p$ zones: one beneath the northern Santorini caldera (labelled B in Fig. 11, $V_p$ < 4.5 km/s), a wider zone under the Kolumbo volcano (labelled C – C', $V_p$ 4.5–5.5 km/s), and a third below Anydros (labelled D, $V_p$ 4.5–5.5 km/s). The three low $V_p$ zones are clearly separated by high $V_p$ volumes, indicating a multi-storage feeder system, and supporting the idea of separate systems feeding Santorini and Kolumbo, in agreement with petrological and geochemical evidence (Klaver et al., 2016).

Beneath Kolumbo volcano, $V_p/V_s$ is high (>1.8) from surface to 10 km depth, revealing the presence of a widespread melt reservoir, which could be the source of the February 2025 dyking event because it closely matches the position of the deflation source identified by Isken et al. (2025) at about 7.5 km depth. At shallower depth (2–5 km, labelled C' in Fig. 11), the high $V_p/V_s$ values are consistent with previous results ($V_p/V_s$ ~ 2.3, Hufstetler et al., 2025), indicating the existence of shallower melt pockets.

Beneath the Santorini caldera, a melt storage zone has been hypothesised at 3–6 km depth (Autumn et al., 2025; Hufstetler et al., 2025; McVey et al., 2019), based only on low $V_p$ anomalies. Despite the large tomographic model uncertainties in the Santorini region (Fig. S15–S16), due to the low level of seismicity compared to that used in previous tomographic models, our model consistently reveals low $V_p$. However, in contrast to the Kolumbo region, the $V_p/V_s$ ratio here is low (<1.75), indicating a greater likelihood of involvement of magmatic fluids rather than melt as a deformation source. The higher $V_p/V_s$ ratio at shallower depth (0–3 km) can be related to liquid water in the hydrothermal system, whereas nothing can be inferred for the deeper structure (> 6 km depth) due to the lack of ray coverage.

The third clear low $V_p$ volume is located underneath Anydros at a depth greater than 8 km, displaying areas with variable $V_p/V_s$ ratios (low, labelled D and high, labelled D' in Fig. 11), and a very low level of seismicity compared to the surrounding zones (refer to Fig. 9B). These features indicate the presence of a magmatic reservoir located beneath, or inside, the Anydros horst (labelled A in Fig. 11). The horst is highlighted in the tomography by the high $V_p$ values ($V_p$ > 6 km/s) compared to the lower values corresponding to the Anydros and Santorini–Anafi grabens ($V_p$ < 5 km/s). Notably, the sudden NE-ward propagation of seismicity that occurred after ebruary 10, 2025 (10 km in less than a day), started from the northeastern boundary of the inferred reservoir at about 10 km depth, possibly suggesting magma movement or fluid release favouring fault activation and propagation.

Near Anydros, the high velocity horst is interrupted by a narrow zone (< 4 km) where $V_p$ is substantially lower (5–6 km/s) and $V_p/V_s$ is slightly higher, corresponding well to the location of a tectonic transverse zone (labelled T in Fig. 11) identified from bathymetric and seismic reflection data (Nomikou et al., 2018) and not resolved in earlier tomography studies (Autumn et al., 2025; Hufstetler et al., 2025). Most events with high non-DC components (>60%) and all events that occurred before February 9, 2025, are concentrated to the southwest (red and green stars in Fig. 11), in the high $V_p$ and low $V_p/V_s$ body, highlighting dilatation and rupture processes within the high-strength Anydros block due to inflation of melt/magmatic fluids. The transverse zone is also associated with overall lower seismicity along the main NE–SW-trending strand (Fig. 9A), and with the area where seismicity shows a temporal migration toward the second strand farther to the NW (Fig. 10F, 10H), indicating a weaker behaviour compared to the surrounding region, able to accommodate and transfer more of the deformation aseismically.

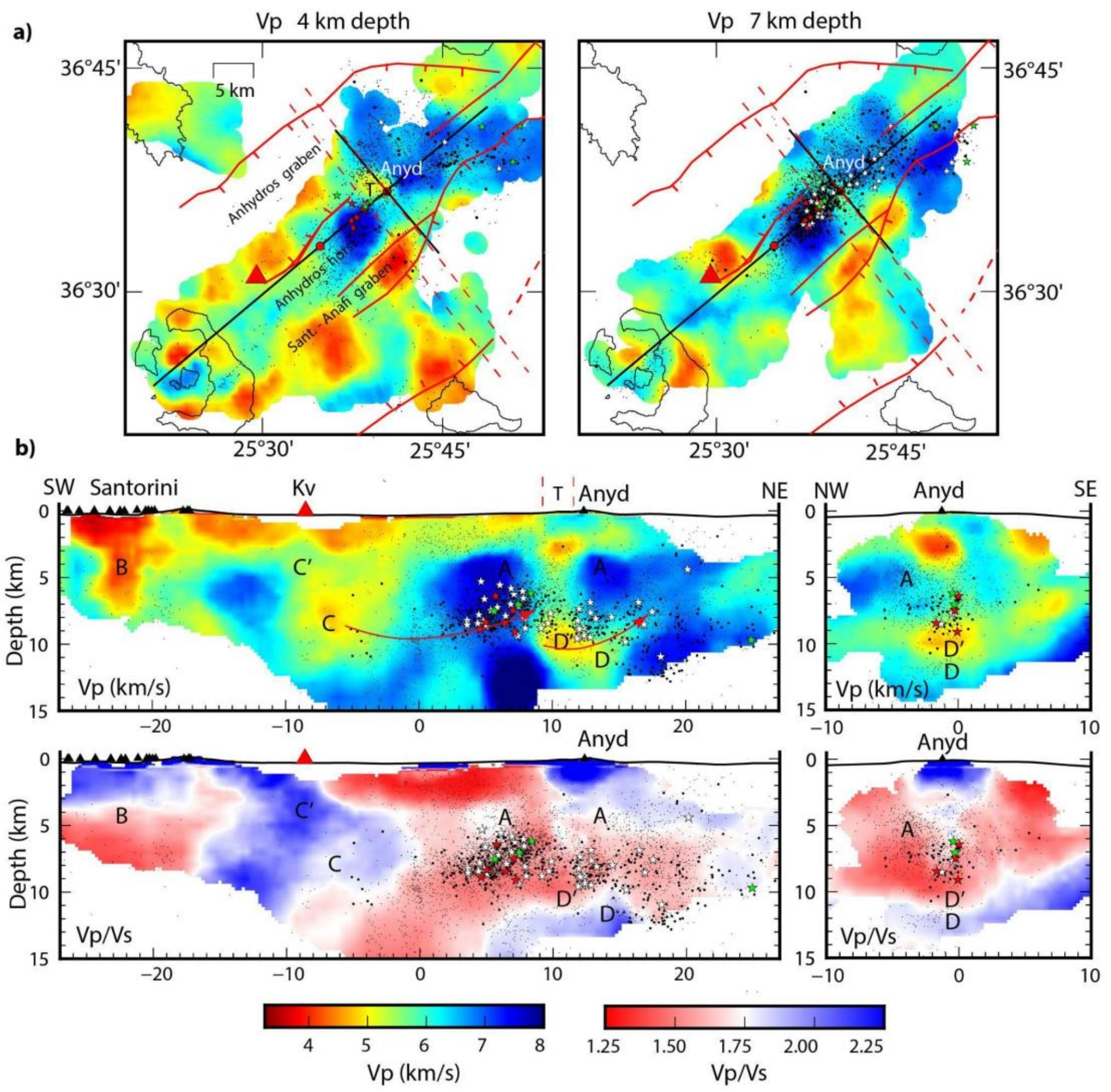

**Figure 11: 3D $V_p$ and $V_p/V_s$ mean posterior models. a) map view at 4 km and 7 km depth. Red lines: main faults (Preine et al., 2022). Dashed red lines: transfer tectonic structure (Nomikou et al., 2018). b) N50°E and N40°W vertical cross-sections. Plotted seismicity within 1 km, sized according to local magnitude (small dots: $M_L < 3$, medium: $3 < M_L < 4$, stars: $M_L > 4$). Red and green stars mark events with high non-DC components (> 60%; Fig. 4C), red: before 09/02/25, green: after 10/02/25. A-D: main anomalies discussed in the text. Kv: Kolumbo volcano. White regions are non-resolved model regions based on ray-coverage and resolution tests (Fig. S17 – S18). Map views of $V_p$ and $V_p/V_s$ models at different depths, together with model uncertainties, are provided in the Supplementary Material (Figs. S14–S16). Red arrows: possible paths for melt/fluid migration. Distance on the x-axis and depth on the y-axis are in km.**

As the State of Emergency was lifted on March 3, 2025 (Timeline, 3 March; Fig. 2), we reached a near real-time high-resolution enhanced DL catalogue of approximately 79,245 events marking a nearly 20 times detection enhancement compared to standard techniques (Timeline *Science*, 10 March; Fig. 2), which illustrated the intensity of the evolving earth processes going far beyond a typical seismic sequence.

## March 24: Implications for the Earth Processes Driving the Unrest

Addressing the urgent questions posed on February 1, 2025, in near real-time posed scientific challenges that were aggravated by practical difficulties, such as the time needed for daily data downloads during critical time windows (>12 hours). This was considerably longer than the local, or cloud-based, data processing via DL workflows (1–2 hours). We shared early results with the Greek academic community on February 2, 2025 (Timeline *Impact*, 2 February; Fig. 2), as well as with Greek and international media in invited interviews, demonstrating the efficiency of DL algorithms under the most pressing conditions. Because public discussion focused on the expected maximum magnitude in what was partially believed to be a non-volcanic sequence (Timeline *Impact*, 3 February; Fig. 2), our team prioritised using the DL catalogue to help interpret the global significance of the complex volcanic-tectonic incident that could be attributed to an intrusion and/or a dyke episode (Timeline *Impact*, 4 February; Fig. 2). The enhancement of the local networks, especially with the deployment of a real-time station on Anydros located above the evolving seismicity, provided critical depth constraints that helped in the assessment of whether an eruption might be imminent. Following the declaration of a state of emergency for Santorini and surrounding islands by the Greek state, the BGS press release (Timeline *Impact*, 7 February; Fig. 2) provided a clearer picture of the ongoing seismicity. Understanding how regional active tectonics would be influenced by the ongoing episode was at the time broadly discussed by team members in the Greek press (Timeline *Impact*, 11 February; Fig. 2). One of the best examples illustrating how DL transforms our understanding in critical time windows came on late February 14, 2025. At that time, the domain expert discussion supported to some extent the presence of banded tremor in Santorini (as reported also in EMSC's social media) (Timeline *Impact*, 14 February; Fig. 2 and Fig. 3, *Is this Volcanic Tremor?*) but enhanced DL detection recognised the intense microseismicity swarm with occurrences approximately every 0.5 sec. Our DL results were openly shared with Greek colleagues, including members of the Santorini crisis risk estimation committee, from the first day of the crisis.

As the seismic crisis decelerated, we turned to volcano geologists to understand how the recent phenomenon matched the rich history of the Santorini volcanic field. Although analogous seismic activity had been observed in volcanic rift settings (e.g., in Iceland, Fischer et al., 2022), the geology in the Santorini–Amorgos region is starkly different because it is not an area of active rifting with frequent basaltic intrusions. Rather, in this environment, it is more likely that deep basaltic intrusions activate discrete mid- to shallow-crustal magma storage regions (typically <15 km), perhaps of more evolved magma as observed at Santorini and Kolumbo (Druitt et al., 2019; Klaver et al., 2016). This could trigger effusive or explosive eruptions (Martin et al., 2006). Although the Kolumbo volcanic chain forms an elongated field of monogenetic submarine cones that extend northeast from Kolumbo, these are each considered the products of separate eruptions over an extended period (Vougioukalakis et al., 2025) and not necessarily related to dyke intrusion. The sources, or storage regions, of the magmas that supplied these cones remain unconstrained. There is no petrological or geochemical evidence for long-distance magma transport laterally in the shallow crust in this region (Francalanci et al., 2026). The NE–SW trail of edifices along the Christiana–Santorini–Kolumbo chain indicates the shallow tectonic control on where magma emerges within the newly recognised Kameni Graben (Preine et al., 2026), most likely from discrete sub-volcanic reservoirs. When high-precision earthquake locations are available near magmatic systems, seismicity often migrates with the movement of pore fluids that are either exsolved from ascending magma or remobilised by heat and geothermal processes (Florez et al., 2025). Thus, we remain open to the possibility that the dyke intruded below the seismicity and the seismicity migration resulted from secondary stress perturbations and fluid migration, as high pressure fluids moved along faults. The 1–2 hour-long seismicity bursts that we observed could have indicated a fault valving process (Shelly et al., 2015) because fluids migrated to areas with lower pore pressure. Large tectonic faults in the area would have been weak surfaces most capable of transporting fluids and most likely to fail when pore pressures increased. Thus, activation of tectonic faults would be expected even if the initial trigger for seismicity was ultimately the ascent or lateral propagation of magma. The time series of the earthquakes, as revealed by the DL catalogue and when compared to global analogues, strongly supported this model. Observations from template matching, MTs and tomography strengthened and refined this interpretation. Because there were few LP earthquakes, no clear shallow tremor, and no immediate signs of heat or degassing at the surface, we were not concerned about an imminent eruption at any point. We infer that magma remained relatively deep.

The public interest remained high, leading to team member presentations sharing preliminary results at the Academy of Athens (Timeline *Impact*, 31 March; Fig. 2). The rapidly migrating activity between January 28 and February 13, 2025, often spanned kilometres within hours to days and coincided with episodic seismic bursts rich in moderate-magnitude events, indicating a strong role for episodic high-pressure magmatic fluid injection, or fluid exsolution, in the dynamics of the sequence. The positive CLVD and ISO components of the moment tensors revealed from the first days of the crisis the likely involvement of magmatic processes or hydrothermal fluids, which may indicate dyking according to Isken et al. (2025) or over pressurisation of hydrothermal fluids on top of the deeper magmatic intrusion near Anydros islet.

Distinguishing which model fits the observations best was a daunting task: in the beginning of the crisis, ground-based geodetic stations existed only on Santorini, and the geometry of the existing seismic network provided no azimuthal coverage to accurately monitor the seismic unrest. On Easter Sunday (April 20, 2025), an event with $M_L = 4.0$ between Santorini and Anafi demonstrated to scientists and locals the importance of clear science communication, especially ahead of the tourist season. As recent work for tsunami estimates (Kalligeris et al., 2025) highlights, "the unusually large displacements in the 1956 earthquake may be repeated in future earthquakes of the region." Multi-hazard scenarios with and without the occurrence of future large mainshocks, an eruption, and associated tsunamis could be estimated, and appropriate mitigation measures could be discussed in advance.

The results and conclusion of this study demonstrate the initial early and evolving state of knowledge regarding the unfolding crisis as it happened, without any form of retrospective analysis. Cloud storage and computing enhanced near real-time analysis, but certain practical problems remained, such as time difference between different locations of research institutes, and other non-stopping academic priorities. These problems inevitably made us wonder whether the gravity of the problem warranted the creation of a dedicated volcanic observatory.

## Perspectives on Dedicated Observatories/Monitoring/Emergency Risk Assessment

Volcano eruption forecasting is a complex task that requires the integration of multiple data streams and nuanced interpretation of seismicity (Acocella et al., 2024; Newhall and Pallister, 2015). Volcano eruption forecasting requires tracking detailed characteristics of seismicity that are lacking in standard earthquake catalogues, such as variations in frequency content, resonant tremor, and potentially transitions when high rates of seismicity merge into tremor-like signals, as was observed in the Santorini swarm. To catch these important, but sometimes subtle, shifts in seismicity, volcano observatories around the world often require scientists to watch the raw data continuously. In this regard, Illsley-Kemp et al. (2021) highlight how enhanced catalogues and other improvements in seismic monitoring could have better informed the decision-making scheme for the Taupo volcano (New Zealand) if they were available at the time. Unlike traditional earthquake catalogues, our DL analysis provided an abundance of information that would otherwise not have been available. Thus, our study demonstrates the potential for DL analyses to inform improved volcano monitoring in the future.

In addition, scientists need more than earthquake catalogues complete to low magnitudes to properly evaluate seismicity near volcanoes. The characterization of the frequency content, the identification of hybrid/LP events, and the temporal evolution of source mechanisms are also critical. These seismic observations would then need to be interpreted in the context of data from other disciplines like geodesy, gas geochemistry, and geological and petrologic history to correctly assess the state of a volcano (Newhall and Pallister, 2015). Given the challenges volcano observatories around the world face and the challenges of having the capacity to perform multiple complex analyses quickly, DL analyses could potentially help inform better management of these situations (X. Tan et al., 2025). The question of whether an earthquake swarm has a purely tectonic or magmatic origin is a frequent problem near volcanoes, and the answer to this question is critical for the effective management of a crisis as it unfolds. The societal and institutional friction generated by these divergent narratives was captured at the time in the SWOT analysis conducted by Synolakis (2025), dated March 10, 2025. This overview evaluates the contemporary scientific communication strategies together with the systematic structural vulnerabilities that affected public understanding at the early days of the crisis. Specifically, the public information ecosystem was impacted by the delay in recognizing the initial seismic unrest, the absence of a dedicated public volcano observatory, and the sparsity of regional seismic, geodetic, and offshore monitoring networks. These technological limitations were exacerbated by entrenched cultural barriers that often impeded cross-institutional data sharing.

Rapid analyses are critical to addressing seismic crises near volcanoes including predictive modelling to guide risk estimation and disaster mitigation. Even the most sophisticated signal processing technology will struggle with sparse seismic monitoring. In island settings, such as the south Aegean volcanic arc, offshore networks would provide lower completeness levels and more accurate event locations. Ideally, ocean-bottom broadband seismometers with telemetry would record and relay critical data streams in emergency situations. Following the unrest, the temporary geodetic network was enhanced, but activity in this region could reawaken at any time, and permanent ground-based geodetic monitoring could help to prepare for that possibility. To improve regional monitoring capabilities, an international collaboration among the Centre National de la Recherche Scientifique (CNRS), Istituto Nazionale di Geofisica e Vulcanologia (INGV), and National Observatory of Athens (NOA) successfully expanded the geodetic network, bringing the total number of operational GNSS stations in the South Cyclades to 32 (21 of them in Santorini) by April 2025. Executing monitoring infrastructure expansion proactively during periods of quiescence rather than implemented as a reactive, post-event response could help ensure that event-based data are accurate and inform decision making. Satellite monitoring capable of tracking water discolouration and sea surface agitation, supported by the European Copernicus programme (Brandl et al., 2020), could provide important updates on ground deformation. Bringing diverse seismic, geodetic, and other geophysical data streams together into a shared physical and digital environment could enhance expert situational awareness during disaster emergencies. Leveraging cloud-based architectures to serve open datasets directly to computational workflows could significantly reduce data transfer bottlenecks.

Recent work (Fountoulakis and Evangelidis, 2025; Isken et al., 2025; Lomax et al., 2025) demonstrates how different methods for seismic catalogue development may result in inconsistencies in earthquake locations that may lead to different interpretations when clarity is urgently needed. Even in retrospective mode, the integration of different datasets ranging from offshore instruments and satellite geodesy (Isken et al., 2025) to ground geodesy (Briole et al., 2025) provided controversial interpretations (Fig. 12). To date (May 2026), the recent deformation of the Santorini caldera (Papazachos et al., 2025), the tectonic and volcanic activity and their role in promoting (Zaccagnino et al., 2025), or not, the 2025 crisis, the role of Kolumbo submarine volcano, the horizontal dyking process (Lomax et al., 2025) or the geometry of the multi-storage magmatic reservoir system, remain unclear. Volcanic systems often exhibit several unrest phases characterised by elevated seismicity without culminating in an eruption; however, these non-eruptive unrests routinely trigger public concern (e.g., Lamb et al., 2024). A dedicated observatory providing reliable, real-time geophysical information could bring together data assimilation, predictive modelling and the human resources required before the next geophysical crisis in the South Aegean Volcanic Arc.

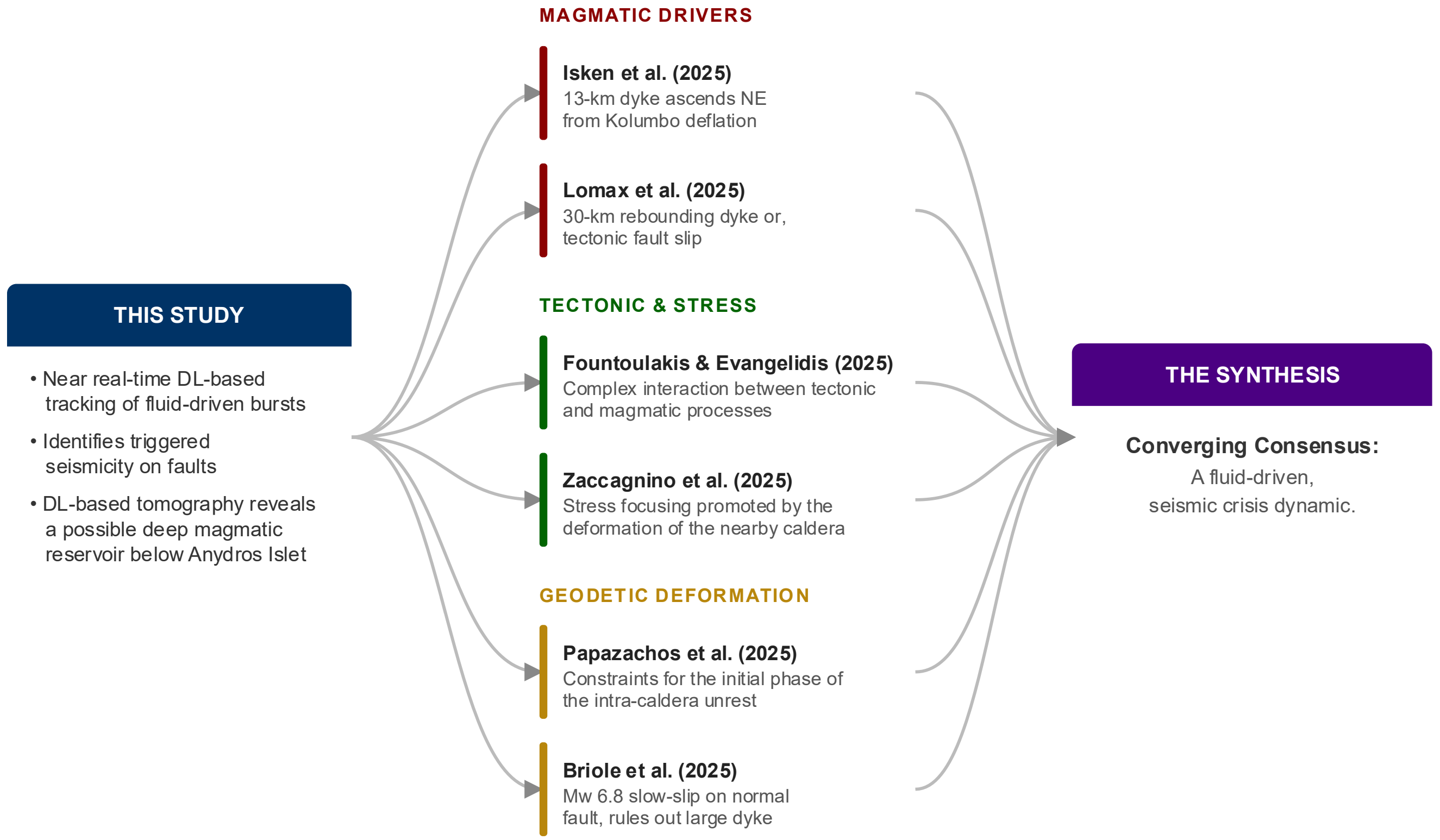

**Figure 12 Schematic of the current understanding behind the 2025 Santorini-Amorgos unrest. A consensus regarding the fluid-driven character of seismicity underlines the importance of our early findings.**

## Author contributions

M. S. led the daily tracking of seismicity, the coordination of the international group, and the manuscript preparation. F. D. worked on seismicity rates analysis, DL earthquake detection, and manuscript writing. X. T. focused on seismicity relocation. P. M. G. estimated moment tensor solutions. F. S. d. U. performed template matching. G. B. guided the interpretation of DL-based seismicity and manuscript writing. G. G. focused on seismic tomography and manuscript writing. W. S. and R. C. aided with local and cloud implementation of DL earthquake detection and data sharing protocols amongst team members. S. P. focused on the interpretation of seismicity, manuscript writing and, together with J. P., led the global analogues discussion, and J. W. focused on LP detection. M. B. contributed with local seismicity interpretation within its seismotectonic setting. D. P. offered critical insights from petrology and local volcanism. C. S. contributed to sharing results with local and committee experts and his experience in understanding tsunami risk. T. P. contributed to manuscript writing. A. G. discussed geodetic data. W. E. contributed with comments on real-time seismicity evolution. B. B. contributed to early analysis of the crisis.

P. P. shared preliminary results on LP detection. G. F., C. C. and W. M. contributed to the interpretation of the unrest.

## Acknowledgements

The work of BGS scientists has been supported by the AI in Seismology International Geoscience Research and Development (IGRD) project. The research was part of the BGS International NC programme 'Geoscience to tackle Global Environmental Challenges', NERC reference NE/X006255/1. The paper is published by permission of the Director of the British Geological Survey. BGS advisories related to the Santorini unrest, submitted to the Foreign, Commonwealth and Development Office (FCDO), were informed by our results. We thank Sotiris Valkaniotis for providing geological data in vector format. The GNSS data are made available by Greek (Hellenic Positioning System HEPOS, NOA GNSS Network NOANET, National Technical University of Athens, National and Kapodistrian University of Athens, University of Patras, Hexagon SmartNet provider, URANUS GNSS Network, JGC Geoinformation Systems S.A.) and international (CNRS, INGV) GNSS networks under the coordination of the Greek Earthquake Planning and Protection Organization (EPPO). Any use of trade, firm, or product names is for descriptive purposes only and does not imply endorsement by the U.S. Government.

## Data availability

Waveforms were accessed daily from the NOA EIDA open data node (https://eida.gein.noa.gr) (Evangelidis et al., 2021). The data used in the search for global analogues are available at https://earthquake.usgs.gov/earthquakes/search/ (ComCat), https://www.isc.ac.uk/ (ISC) (International Seismological Centre, 2026), (SIL) (Bodvarsson et al., 1996), https://www.jma.go.jp/jma/index.html (JMA), https://www.ign.es/web/ign/portal/sis-catalogo-terremotos (IGN) and https://volcano.si.edu/ (GVP) (Global Volcanism Program and Venzke, 2025). The earthquake catalogues produced in this study are available on Zenodo (Segou et al., 2026).

# Enhanced Seismicity Monitoring in the Rapid Scientific Response to the 2025 Santorini Crisis – Supplementary Material

Margarita Segou*[1], Foteini Dervisi[1,2], Xing Tan[3], Rajat Choudhary[1], Patricia Martínez-Garzón[4,5], Francesco Scotto di Uccio[6], Gregory Beroza[3], Genny Giacomuzzi[7], Claudio Chiarabba[7], Wayne Shelley[1], Stephanie Prejean[8], Jeremy Pesicek[9], John J. Wellik[9], Marco Bohnhoff[4], David Pyle[10], Costas Synolakis[11,12], Tom Parsons[13], Athanassios Ganas[14], William Ellsworth[3], Brian Baptie[1], Gaetano Festa[6,7], Piero Poli[15], and Warner Marzocchi[16]

[1] The Lyell Centre, British Geological Survey, Edinburgh, UK.
[2] School of GeoSciences, The University of Edinburgh, Edinburgh, UK.
[3] Department of Geophysics, Stanford University, Stanford, CA, USA.
[4] GFZ Helmholtz Centre for Geosciences, Potsdam, Germany.
[5] RWTH Aachen University, Aachen, Germany.
[6] Dipartimento di Fisica Ettore Pancini, Università degli Studi di Napoli Federico II, Naples, Italy.
[7] Istituto Nazionale di Geofisica e Vulcanologia (INGV), Rome, Italy.
[8] U.S. Geological Survey, Volcano Disaster Assistance Program, Anchorage, AK, USA.
[9] U.S. Geological Survey, Volcano Disaster Assistance Program, Vancouver, WA, USA.
[10] Department of Earth Sciences, University of Oxford, Oxford, UK.
[11] Department of Civil and Environmental Engineering, University of Southern California, Los Angeles, CA, USA.
[12] Division of Natural Sciences, Academy of Athens, Athens, Greece.
[13] U.S. Geological Survey, Pacific Coastal and Marine Science Center, Moffett Field, CA, USA.
[14] Institute of Geodynamics, National Observatory of Athens, Athens, Greece.
[15] Dipartimento di Geoscienze, Università degli Studi di Padova, Padua, Italy.
[16] Dipartimento di Scienze della Terra, dell' Ambiente e delle Risorse, Università degli Studi di Napoli Federico II, Naples, Italy.

*msegou@bgs.ac.uk [corresponding author]

## Overview of Methods

We analysed permanent and temporary stations within 100 km distance from Santorini Island. At the mid of February, 24 continuous seismic stations were available through the National Observatory of Athens (NOA) data node (Evangelidis et al., 2021) of the European Integrated Data Archive (EIDA), accessed daily. Station ANYD on Anydros Island situated on top of the evolving seismicity became operational on February 3, 2025, and station IOSI became available on Ios Island on February 13, 2025.

Here, we summarise the methodology that was followed. The analysis presented is based on the QuakeFlow (Zhu et al., 2023) deep learning (DL) workflow with further information presented in the *Machine Learning Event Detection* section below. Enhancement of the DL catalogue was done using EQCorrscan (Chamberlain et al., 2017), with details in the *Catalogue Enhancement: Template Matching* section and moment tensor calculation following the FOCI approach (Kwiatek et al., 2016) is detailed in the *Moment Tensor* section. The Frequency Index (FI) (Buurman & West, 2010) was used for frequency content characterisation focusing on long-period-enriched earthquakes, and TomoDD (Zhang & Thurber, 2003) was used for relocation of seismicity, incorporating a 1D velocity model inverted with VELEST (Kissling et al., 1994, 1995). A non-linear Bayesian trans-dimensional inversion method (Giacomuzzi et al., 2024; Piana

Agostinetti et al., 2015) was used to conduct seismic tomography, and is described in detail in the *Seismic Tomography* paragraph.

## Machine Learning Event Detection

We used the QuakeFlow (Zhu et al., 2023) workflow, including PhaseNet, a DL phase picker (Zhu & Beroza, 2019), and GaMMA (Zhu et al., 2022), an unsupervised earthquake associator based on a Bayesian Gaussian Mixture Model. We used the velocity model from Bohnhoff et al. (2006) in GaMMA to associate picks into seismic events, and determine their origin times and initial locations. We set the minimum number of picks per event to 10 recorded over 5 seismic stations without imposing additional constraints on the minimum number of P and S picks for each event. GaMMA successfully associated 451,860 of the total 707,190 P picks and 513,314 of the total 796,610 S picks into 41,873 events for the monitoring period (January 25 through March 3, 2025).

## Catalogue Enhancement: Template Matching

We extended the machine learning catalogue by integrating the template matching detection technique EQCorrscan (Chamberlain et al., 2017), using the earthquakes individuated by PhaseNet + GaMMA to form a template set. The adoption of machine learning for a starting catalogue has been shown to improve the detection resolution of template matching compared to the selection of the existing standard catalogue (Scotto di Uccio et al., 2023), due to the higher number of identified earthquakes and population in the low magnitude domain. We trimmed template events 1.5s around the PhaseNet P and S pick for each earthquake, included 0.15s of pre-pick, downsampled the waveforms to 25 Hz and applied a 2–9 Hz bandpass filter (Chamberlain et al., 2017; Scotto di Uccio et al., 2023; Vuan et al., 2018). We used each template to scan the continuous records, which underwent the same preprocessing operation. We declared an event when the stacked similarity among the stations overcame 10 times the median absolute deviation (MAD) of the cross-correlation coefficients. For each event detected by template matching, we performed cross-correlation phase picking with the most similar template, and imposed a minimum, absolute, cross-correlation coefficient of 0.7 at individual stations for determining the phase arrival times. Template matching detection algorithms leverage stacked cross-correlations across the entire network (Chamberlain et al., 2017; Vuan et al., 2018), resulting in a global similarity value to which stations with both high and low cross-correlation values contribute (Scotto di Uccio et al., 2024). Therefore, low magnitude earthquakes typically report phase arrival times only at the stations closest to the epicentres, hindering the possibility of accurately retrieving the hypocentral coordinates. For reference events, we assign the same hypocentral coordinates as the most similar template in the PhaseNet catalogue. Moreover, we evaluated local magnitude for the events in the template matching catalogue through the amplitude ratio of 1 Hz high-pass filtered horizontal displacement, assuming colocation with the corresponding master earthquakes:

$$Ml_{reference} = Ml_{template} + log_{10}\left(\frac{A_{reference}}{A_{template}}\right). \quad (1)$$

## Analysis of the Magnitude – Frequency Distribution

We investigated the magnitude – frequency distributions (MFDs) for the derived seismic catalogues using ZMAP (Wiemer, 2001). We used the maximum curvature method (Wiemer & Wyss, 2000; Wyss et al., 1999) and estimated slope changes in the MFDs, which we interpret as the most frequently observed magnitude.

## Seismicity Rate Change Quantification

To quantify the hourly seismicity rate change, we calculated the $\beta$ statistic (Matthews and Reasenberg, 1988) in rolling one-hour intervals as

$$\beta = \frac{N_a - N_b \frac{\Delta t_a}{\Delta t_b}}{\sqrt{N_b \frac{\Delta t_a}{\Delta t_b}}} \quad (2)$$

with $N_a$ representing the number of events within $\Delta t_a$, the time window of interest, and $N_b$ representing the number of events within $\Delta t_b$, the reference time window. We used the previous hour as the reference time window and the next hour as the time window of interest in rolling intervals, thus identifying increases and decreases in seismicity rates between consecutive hours. Values of $\beta > 2$ indicate statistically significant increases in the seismicity rate (Hill & Prejean, 2015; Marsan & Wyss, 2011).

## Moment Tensor Calculation

For the calculation of moment tensors, we extracted waveform data from seismic stations located at distances up to 150 km from the epicentre. We selected seismic events in the periods between February 3 (the start of operations of ANYD station) and February 27, 2025. We focused on the characterisation of the largest seismic events of the sequence (M ≥ 3.8, Fig. S1). Hence, we utilised the seismic catalogue from NOA as a reference catalogue for locations ("standard catalogue"). This yielded an initial set of 160 seismic events.

The waveforms from these seismic events were detrended and filtered with a second-order Butterworth high-pass filter of 0.5 Hz to remove undesired low-frequency noise. We manually picked the duration of the first-motion P-wave on the vertical components of ground displacement seismograms. From this duration, we calculated the area beneath the first P-pulse, which is proportional to the seismic moment. We initially imposed a minimum of 8 phases to further proceed. Due to the uneven distribution of the stations, with a large station density towards the south-west of the sequence (Santorini Island), we only calculated moment tensors for earthquakes for which the azimuthal gap was smaller than 100° and the station ANYD (directly on top of the sequence) could be picked. We calculated take-off angles by using a local 1D velocity model (Bohnhoff et al., 2006).

Many moment tensor inversion techniques rely on fitting synthetic seismograms to the waveform recordings of specified wave trains (Dreger, 2015; Sokos & Zahradnik, 2008). However, fitting synthetic waveforms is problematic for small earthquakes due to limited data availability, noise and unmodelled Earth structure at high frequencies, and the increased influence of attenuation. Comparison of different moment tensor inversions indicates that P-wave amplitudes can be successfully used to estimate full moment tensors of small-magnitude events (Bentz et al., 2018; Guilhem et al., 2014).

We calculate moment tensors following the FOCI approach (Kwiatek et al., 2016). This technique uses the area below the first-motion P-pulse, estimates the full, deviatoric and double-constrained moment tensor solutions. The resulting seismic MT is decomposed into its isotropic (ISO) and deviatoric part, including a compensated linear vector dipole (CLVD) and a double couple (DC) component, following the convention introduced by (Knopoff & Randall, 1970):

$$M = M_{ISO} + M_{CLVD} + M_{DC}, \quad (3)$$

where

$$M = \frac{1}{3}tr(M)\begin{pmatrix}1 & 0 & 0\\0 & 1 & 0\\0 & 0 & 1\end{pmatrix} + |\epsilon| m^*_{max}\begin{pmatrix}-1 & 0 & 0\\0 & -1 & 0\\0 & 0 & 2\end{pmatrix} + (1-2|\epsilon|)m^*_{max}\begin{pmatrix}-1 & 0 & 0\\0 & 0 & 0\\0 & 0 & 1\end{pmatrix}, (4)$$

in which $\epsilon = \frac{-m^*_{min}}{|m^*_{max}|}$, and $m^*_{min}$ and $m^*_{max}$ are the eigenvalues of the deviatoric moment tensor with the minimum and maximum absolute values respectively (Vavryčuk, 2001). The percentages of the individual MT components can be calculated following two different definitions of the scalar seismic moment (compare eq 13 with eq 14 in Vavryčuk, 2015). The ISO part describes changes in the volume in the seismic source region. The deviatoric part (CLVD and DC) is used to estimate the orientation of tensional, compressional and null-axis directions (P, T, and B axes), fault-plane orientation, and the slip vector. Furthermore, the covariance matrix of the MT components is also provided. The faulting type is categorised into either strike slip, normal, or thrust faulting, depending on plunges of P, T, and B axes of the resolved MT.

The deviatoric solution decomposes the moment tensor into double-couple and CLVD components. The deviatoric solution is obtained by assuming no volumetric change in the resolved MT during linear inversion. The DC constrained MT is resolved by further imposing the determinant of the seismic MT to be zero. The latter constraint makes the MT inversion scheme nonlinear, and we use Lagrange multipliers (Oncescu, 1986) to determine the DC MT.

Uncertainties of the estimated MTs can be estimated through the normalised root mean square (RMS) error between theoretical and estimated amplitudes (Stierle et al., 2014), following the formula:

$$RMS = \sqrt{\frac{\sum_{i=1}^{N}\left(u_i^{obs} - u_i^{th}\right)^2}{\sum_{i=1}^{N}\left(u_i^{obs}\right)^2}} \quad (5)$$

To further evaluate the variability of the moment tensors, we applied a "Jacknife technique" to the stations used in the inversion. In addition to the "best solution" calculated from all available stations, we also performed the same moment tensor inversion by removing each one of the stations one at a time. This technique allows us to evaluate the dependence of the moment tensor on individual stations and how well constrained they are.

We used the previous constraints to obtain moment tensors for 128 seismic events. All moment tensors were visually reviewed. Three moment tensors displaying remarkable polarity mismatch were eliminated, leaving 125 moment tensors for further analysis. The RMS distributions of the three individual inversions are shown in Fig. S7a. Inverted moment magnitudes typically show good agreement and ± 0.1 units of magnitude difference with respect to those reported by NOA (Fig S7b). The number of readings used to calculate moment tensors varies between 8 and 22, with a median value of 18 readings (Fig. S7c). Examples of representative moment tensors and their stability with respect to the type of inversion and Jacknife test are shown in Fig. S8. These solutions provide information about the deviatoric state of stress of the region.

## Seismic Tomography

We used a fully non-linear Bayesian, trans-dimensional, and hierarchical approach to estimate the velocity models by Local Earthquake Tomography (Bodin et al., 2012; Giacomuzzi et al., 2024; Piana Agostinetti et al., 2015). This method was first used to relocate the DL-derived catalogue in the 1D velocity model (Bohnhoff et al., 2006), without allowing velocity parameters to change, and aiming to select the events that can be more accurately relocated. We selected about 13k events (161,122 P-phases, 147,403 S-phases, Table S2), having relocation errors (standard deviations obtained from the posterior density distribution) lower than 1.5 km, RMS < 0.25s and a minimum number of 5 P- and 3 S-phases. Then, the selected events were used to perform the 3D tomography, allowing changes in both earthquake parameters and velocity structure.

Using this approach for the tomographic problem, the solution is a Posterior Probability Distribution (PPD), defined by the Bayes' theorem:

$PPD = p(\boldsymbol{m}|\boldsymbol{d}_{obs}) \propto p(\boldsymbol{d}_{obs}|\boldsymbol{m})p(\boldsymbol{m})$ (6)

The Earth's structure is parametrised by an ensemble of Voronoi cells, whose number, size, $V_p$ and $V_p/V_s$ values, and position are considered as unknowns. The prior probability distributions $p(\boldsymbol{m})$ are summarised in Table S1. The likelihood function $p(\boldsymbol{d}_{obs}|\boldsymbol{m})$ is based on a simple least squares misfit function $\varphi(\boldsymbol{m})$:

$\varphi(\boldsymbol{m}) = \left(\boldsymbol{d}_{obs} - g(\boldsymbol{m})\right)^T \boldsymbol{C_d}^{-1}\left(\boldsymbol{d}_{obs} - g(\boldsymbol{m})\right) \Rightarrow L = p(\boldsymbol{d}_{obs}|\boldsymbol{m}) = \frac{1}{(2\pi^N|\boldsymbol{C_d}|)^{1/2}} e^{-\frac{\varphi(m)}{2}}$ (7)

where $g(\boldsymbol{m})$ is the calculated arrival time data, $N$ is the dimension of vector $\boldsymbol{d}_{obs}$, and $\boldsymbol{C_d}$ is the data covariance matrix, which here is assumed to be diagonal.

Uncertainties in phase picks were estimated by grouping them into four classes (Table S2), based on the peak classification probability estimated by PhaseNet ($p$). We assigned weight 0 for $p \geq 0.95$, weight 1 for $0.8 \leq p < 0.95$, weight 2 for $0.6 \leq p < 0.8$ and weight 3 for $p < 0.6$. We adopted a hierarchical approach (Bodin et al., 2012) that treats the overall level of data noise as an unknown parameter to be estimated. Initial pick uncertainties ($\sigma$) associated with the different classes are 0.08 seconds (s), 0.12 s, 0.2 s and 0.6 s. We considered two different noise levels for P-wave and S-wave data, which rescale the picking uncertainties ($\varepsilon_i = 10(n + log(\sigma_i))$; $\varepsilon_i$ is the estimated data noise for the $i$-th arrival times (s), $n$ the unknown noise level, $\sigma_i$ the starting picking uncertainty). Observed data are considered independent, with uncorrelated noise. Both the P- and S-phases noise parameters estimated *a posteriori* are very close to zero (0.02 and $-0.05$), indicating that a reliable data weighting scheme was used.

The PPD is evaluated with the Reversible Jump Markov Chain Monte Carlo (RJ-MCMC) sampling algorithm (Mosegaard & Tarantola, 1995). The chain starts with a model randomly sampled from the prior distributions and models are generated according to a sampling recipe (Table S3) and accepted/rejected based on the Metropolis-Hastings rule (Green, 2003). Chain convergence is considered achieved when the noise level parameters stop decreasing (Fig. S19a). After reaching convergence, one sample every 1k from the last 250k generated models is collected. To ensure an optimal sampling of the PPD, we run 20 Markov chains, each starting from a different random point of the prior distributions. The final PPD is thus formed by an ensemble of 5k collected models. The calculation was performed on the Mercalli INGV cluster, using 640 cores for approximately 120 hours.

Theoretical arrival times were computed using the pseudo-bending ray-tracing in a 3D velocity model defined by a Cartesian grid of nodes. Voronoi tessellations are rasterised onto the grid by associating, at each node, the $V_p$ and $V_p/V_s$ values of the Voronoi cell that includes the grid node itself (grid interspacing amounts to 0.5 km in the central part of the model). The same grid node is used to compute frequency distributions for $V_p$ and $V_p/V_s$ (Fig. S16b). Mean $V_p$ and $V_p/V_s$ values computed from the frequency distributions, along with the distribution of model errors, are represented using standard tomographic horizontal and vertical cross-sections (Figs. 11, S14 – S15). We estimate model errors using the absolute deviation ($\sigma = \frac{1}{N}\sum_{i=1}^{N}|x_i - m|$), because posterior frequency distributions are not always unimodal. Similarly, mean values and associated uncertainties for event coordinates were derived from their respective posterior frequency distributions.

The analysis of model uncertainties does not provide information about the minimum size of detectable seismic anomalies or the possible smearing of anomalies into the surrounding volume. We performed a checkerboard resolution test (Fig. S18) with the same global search parameters as those used for the observational data inversion. The station and event distributions are identical, and random noise was added to synthetic arrival times to simulate observational data noise. Event locations were treated as unknown parameters like in the real data inversion. The input anomalies have an amplitude of 20% relative to the background 1D model, with opposite signs for $V_p$ and $V_p/V_s$, and a spatial dimension of 4–5 km in all three directions. This can be considered the minimum size of interpretable velocity anomalies in the Santorini area (down to 7 km depth) and beneath the region with the highest seismicity (down to 13 km depth). The greatest smearing, mainly in the SW–NE direction, is observed in the Kolumbo region, where the minimum size of recoverable velocity structures is likely greater. The results also highlight the comparable resolution between the $V_p$ and $V_p/V_s$ models.

Final RMS values, computed using the mean posterior velocity models and mean posterior earthquake locations, are 0.11 s and 0.15 s, for P- and S-residuals. Compared to the RMS computed in the 1D velocity model, this represents an improvement of 25% for P-residuals and 37% for S-residuals (Fig. S19b).

## Figures

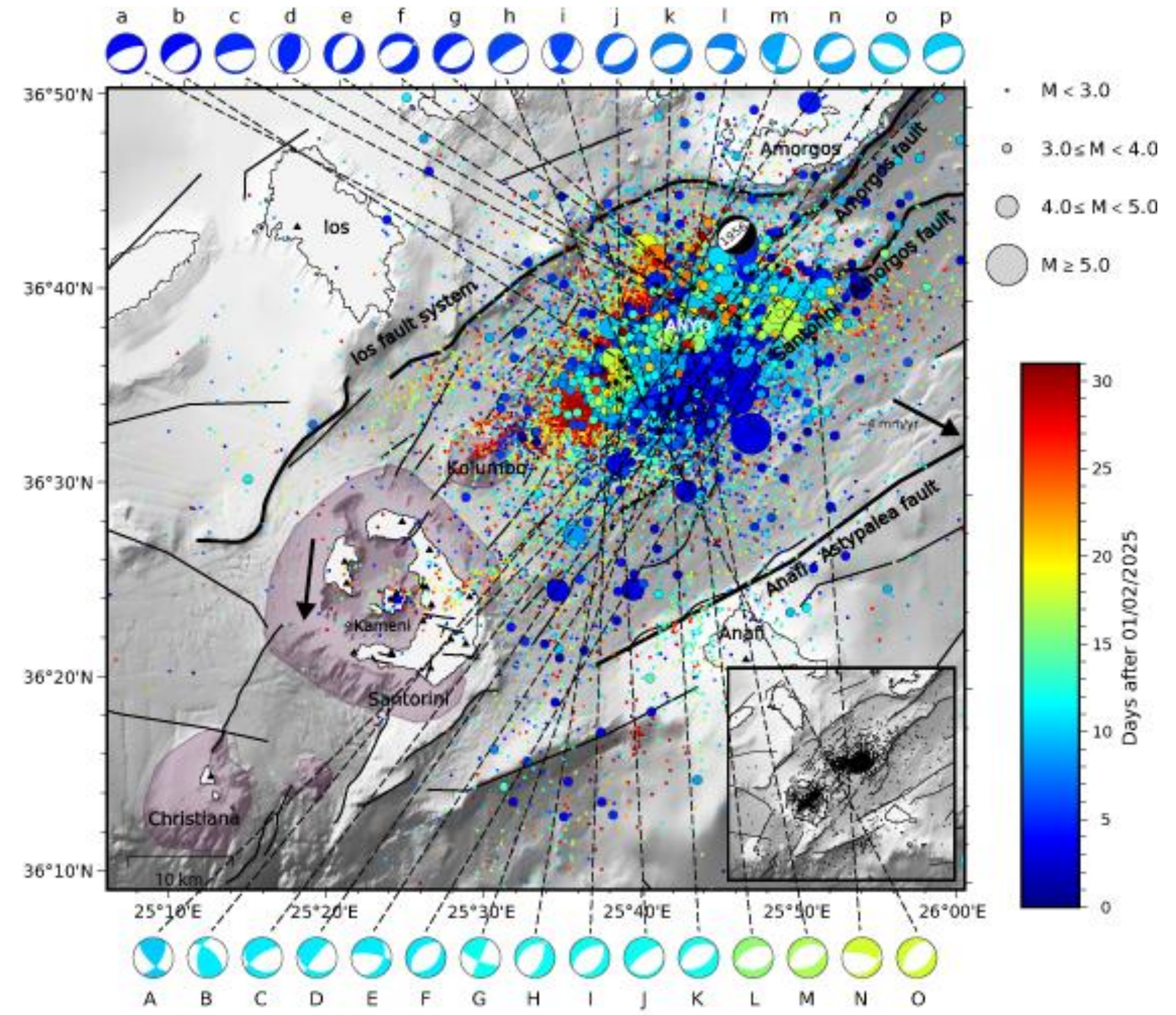

**Figure S1.** Overview of the 2025 sequence. Santorini volcano and the submarine Kolumbo volcanic chain, active faults and high-resolution bathymetry, as taken from the National Observatory of Athens (Ganas, 2025) and the European Marine Observation and Data Network (EMODnet Bathymetry Consortium, 2024), respectively. Black dots in the inset denote the seismicity observed during June 1, 2024–January 31st, 2025, whereas events that occurred from February 1 through March 3, 2025, are colour-coded by time. Seismicity within the aforementioned periods results from the implementation of the QuakeFlow (Zhu et al., 2023) DL workflow of this study. Double-couple moment tensors of events from this study with magnitude M > 4, colour-coded by time since February 1, 2025, mainly indicated normal faulting kinematics consistent with the main tectonic fabric of the region, but diverse combinations of reverse and strike-slip ruptures are also present. Geodetic vectors show the extensional strain field (McClusky et al., 2000). Seismic stations operating by the end of the seismic crisis are shown with black triangles. The black and white focal mechanism is that of the 1956 Amorgos mainshock (Brüstle et al., 2014).

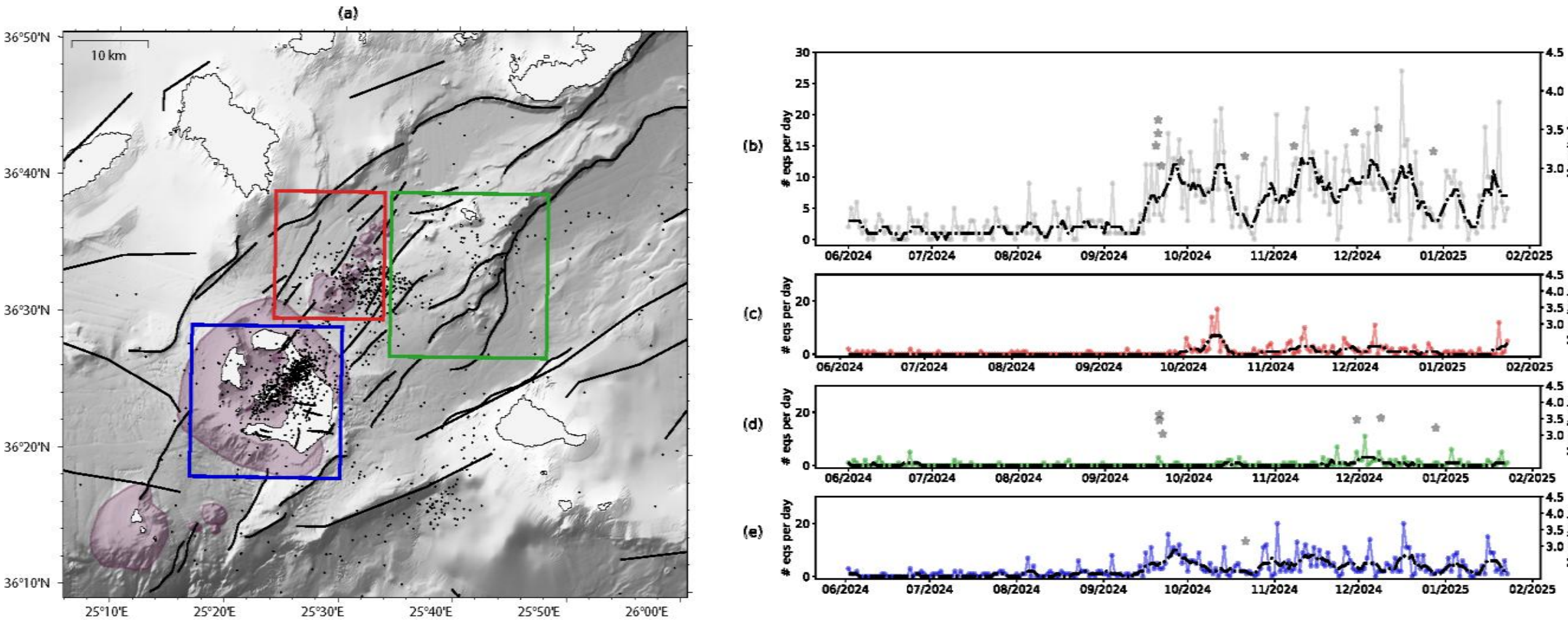

**Figure S2.** The pre-seismic unrest earthquake catalogue enhanced by DL detection. Increased microseismicity near Anydros and Santorini initiates after mid-September with <10 events per day. During June 1, 2024–January 28, 2025, seismicity remained elevated at varying rates across the three reservoirs. Santorini caldera did not present any triggered seismicity during the February 2025 unrest even when the DL catalogue was further enhanced by template matching (Fig. 8). During the seismic crisis, seismicity remains within the Anydros ridge with no triggered seismicity (Fig. 9G–H) at Kolumbo volcano. This shows that interaction patterns between the three magmatic reservoirs, stress-mediated or otherwise, vary across time.

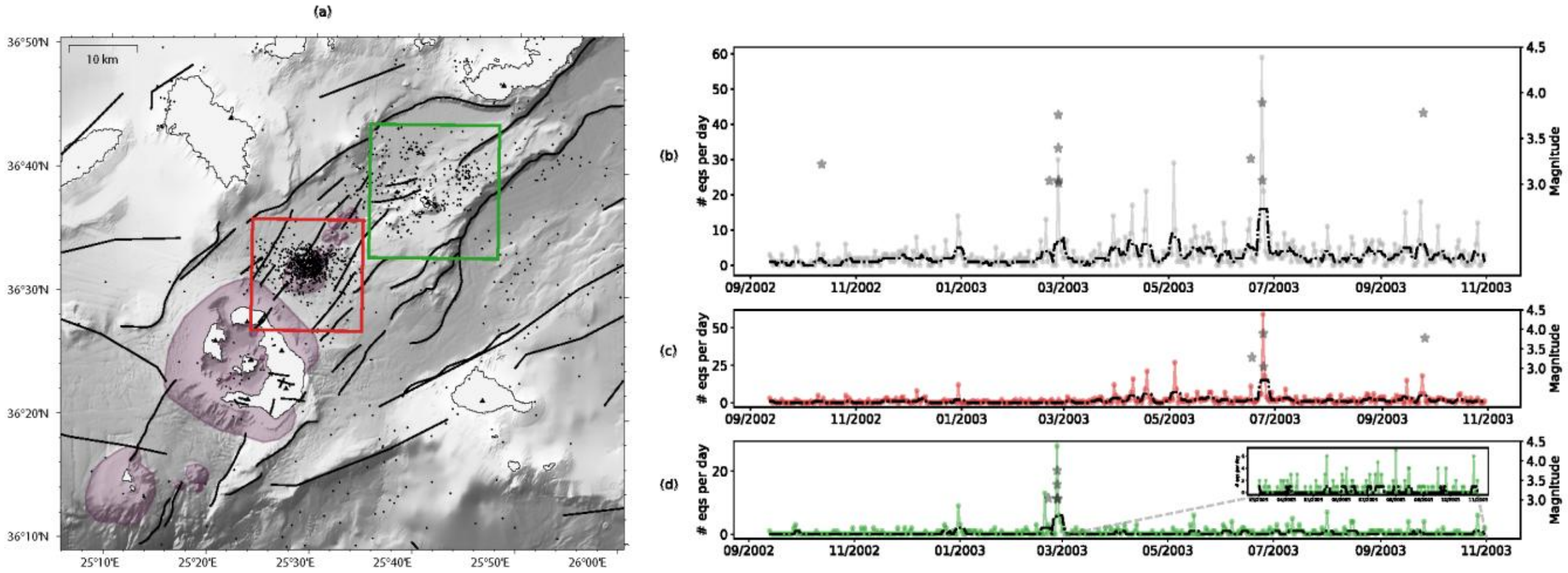

**Figure S3.** The 2002–2003 seismic crisis between Santorini and Amorgos enhanced by DL detection. Seismicity near Anydros presented similar burst-like features with maximum magnitude reaching $M_L = 3.8$. The clearer image provided by DL detection reveals that clustered seismicity in Kolumbo submarine volcano (red box) was triggered shortly after the March 2003 seismicity intensification near Anydros and lasted up to 4 months. The DL catalogue of the 2002–2003 seismic crisis of this study provides the first line of evidence that the tectonic horst of Anydros islet and the Kolumbo volcano interact, but the conditions and physical mechanisms are not straightforward. Bohnhoff et al. (2006) has improved monitoring using the island-based deployment (temporary stations shown as triangles) data reaching 2,175 events instead of 127 events in the standard catalogue (September 2002–July 2004), whereas this study finds 3,109 (September 2002 – October 2003).

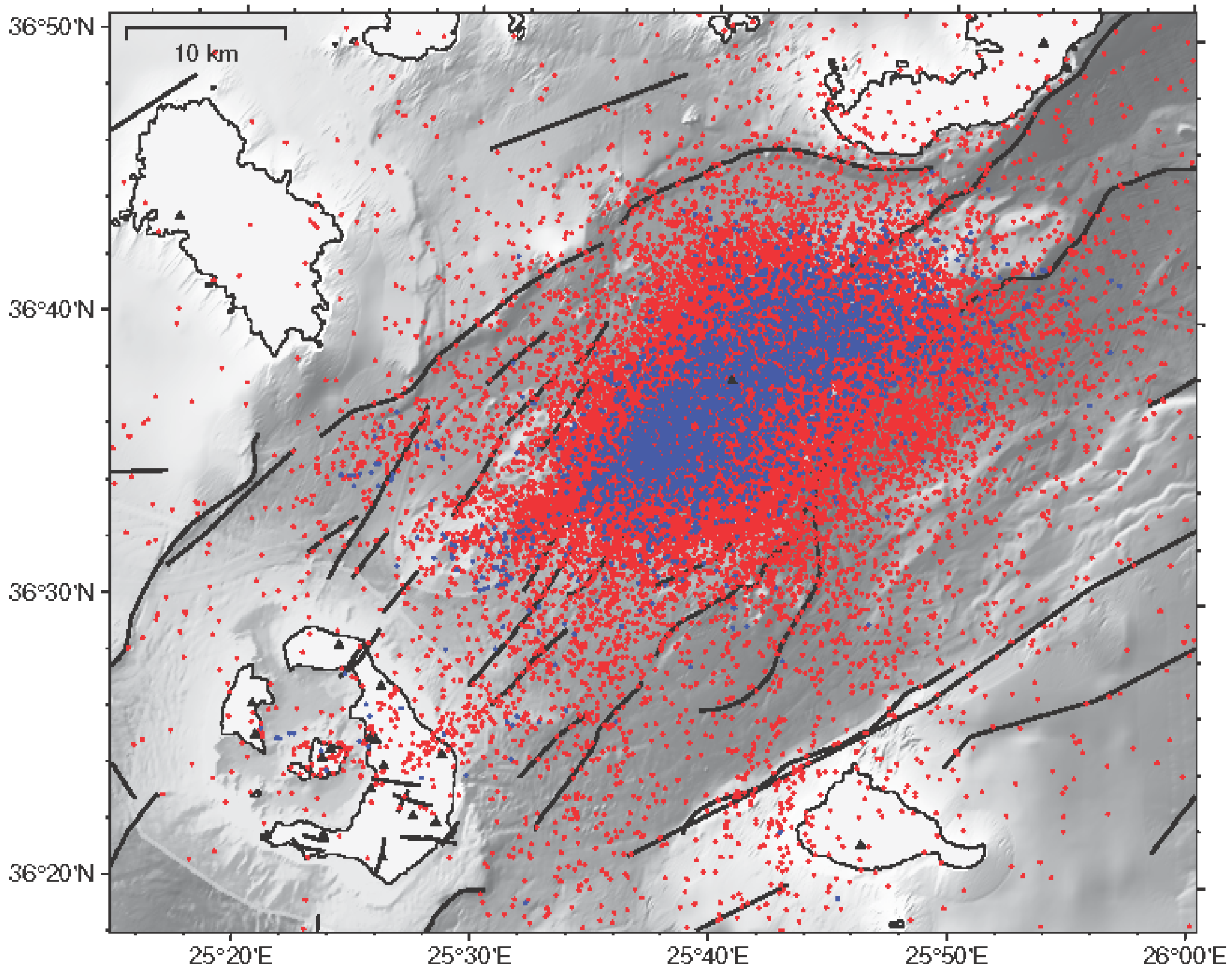

**Figure S4.** Seismicity map of DL-based detection (red) versus the standard catalogue from the National Observatory of Athens (blue). An earlier version of this figure was included in the press release of the British Geological Survey, February 7, 2025 (BGS Press, 2025).

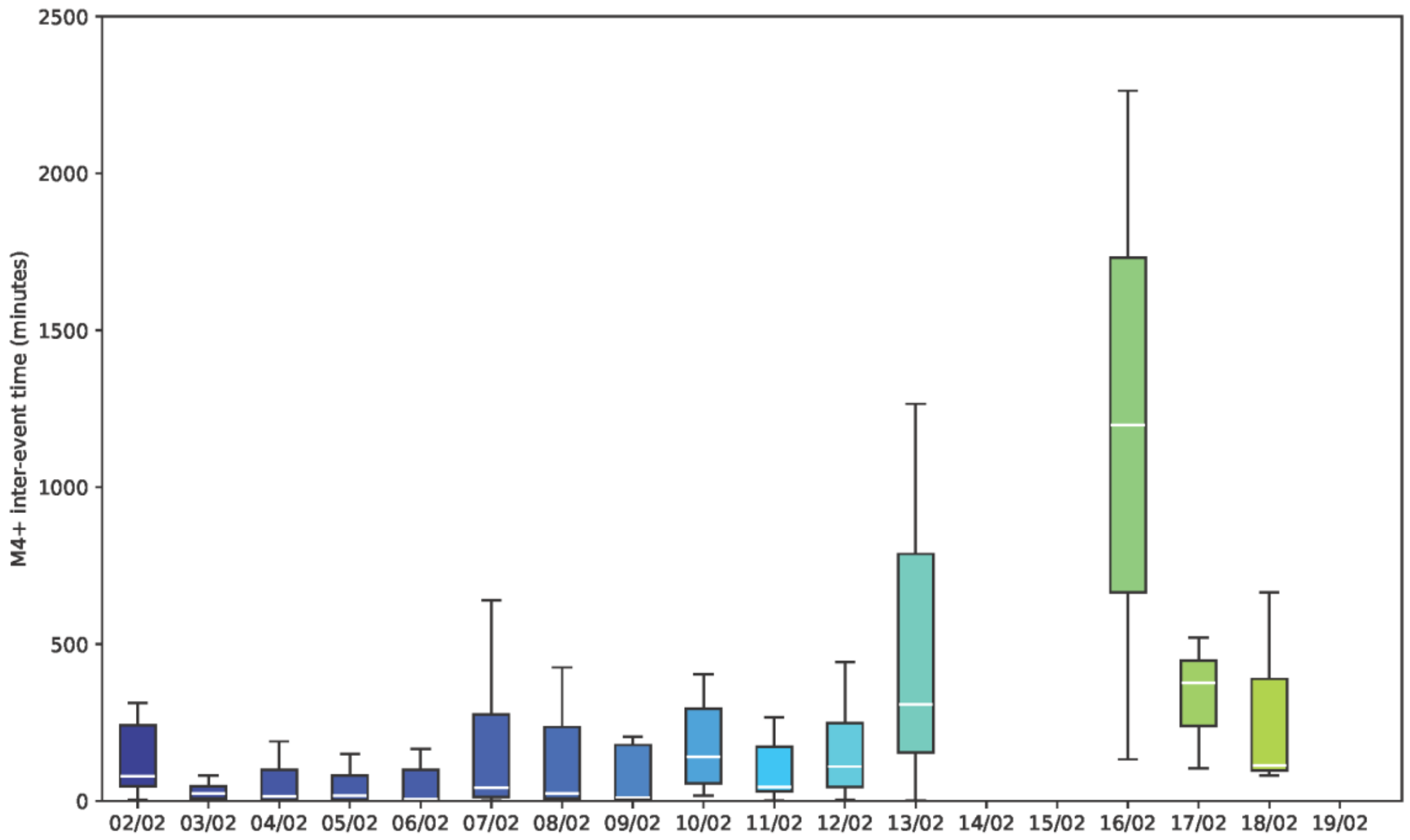

**Figure S5.** Inter-event times between moderate to high-magnitude earthquakes (M ≥ 4) over time.

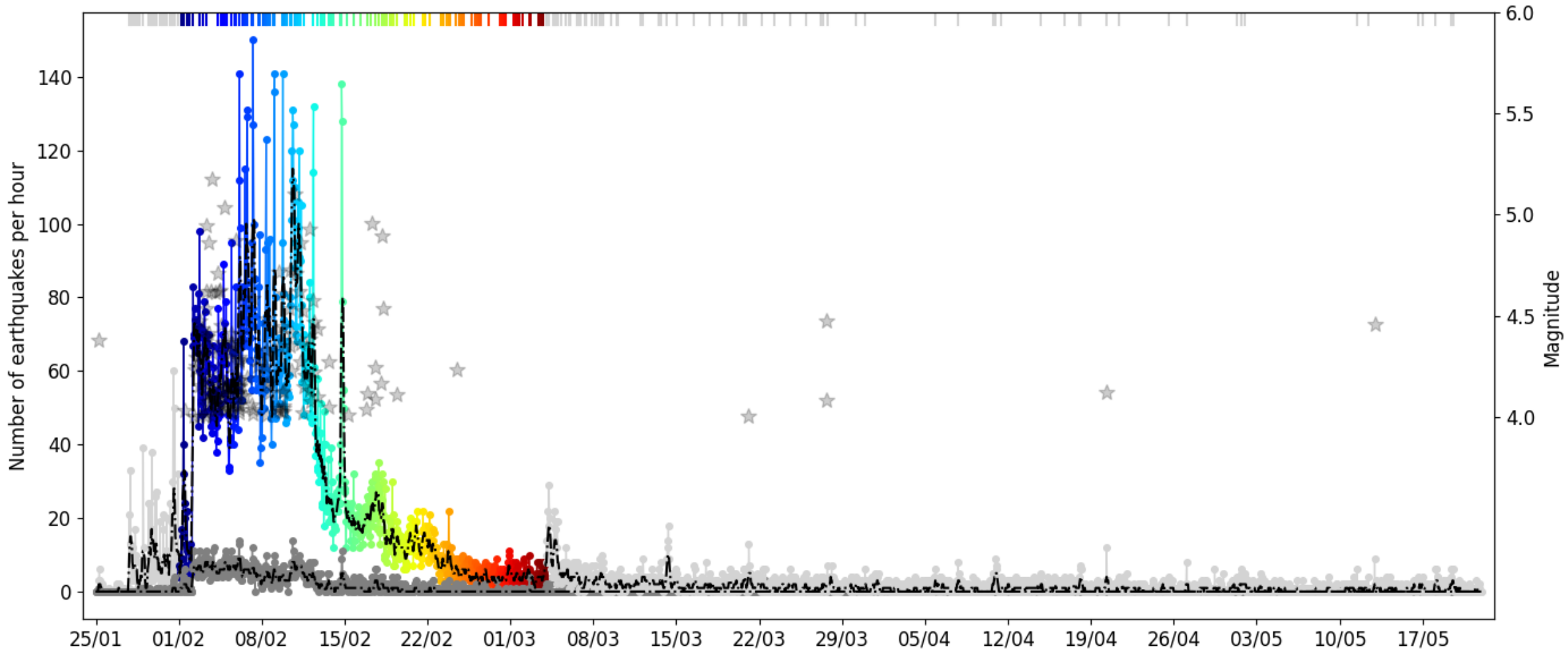

**Figure S6.** Daily tracking of Santorini seismicity using the DL workflow for January 25–May 20, 2025. The rates calculated from the DL catalogue are shown in colour for the crisis dates (February 1–March 3, 2025) and light grey for before and after the crisis, whereas the rates calculated from the standard catalogue are shown in dark grey. The vertical lines on top denote times when the $\beta$ statistic was over 2, indicating significant hourly seismicity rate increases.

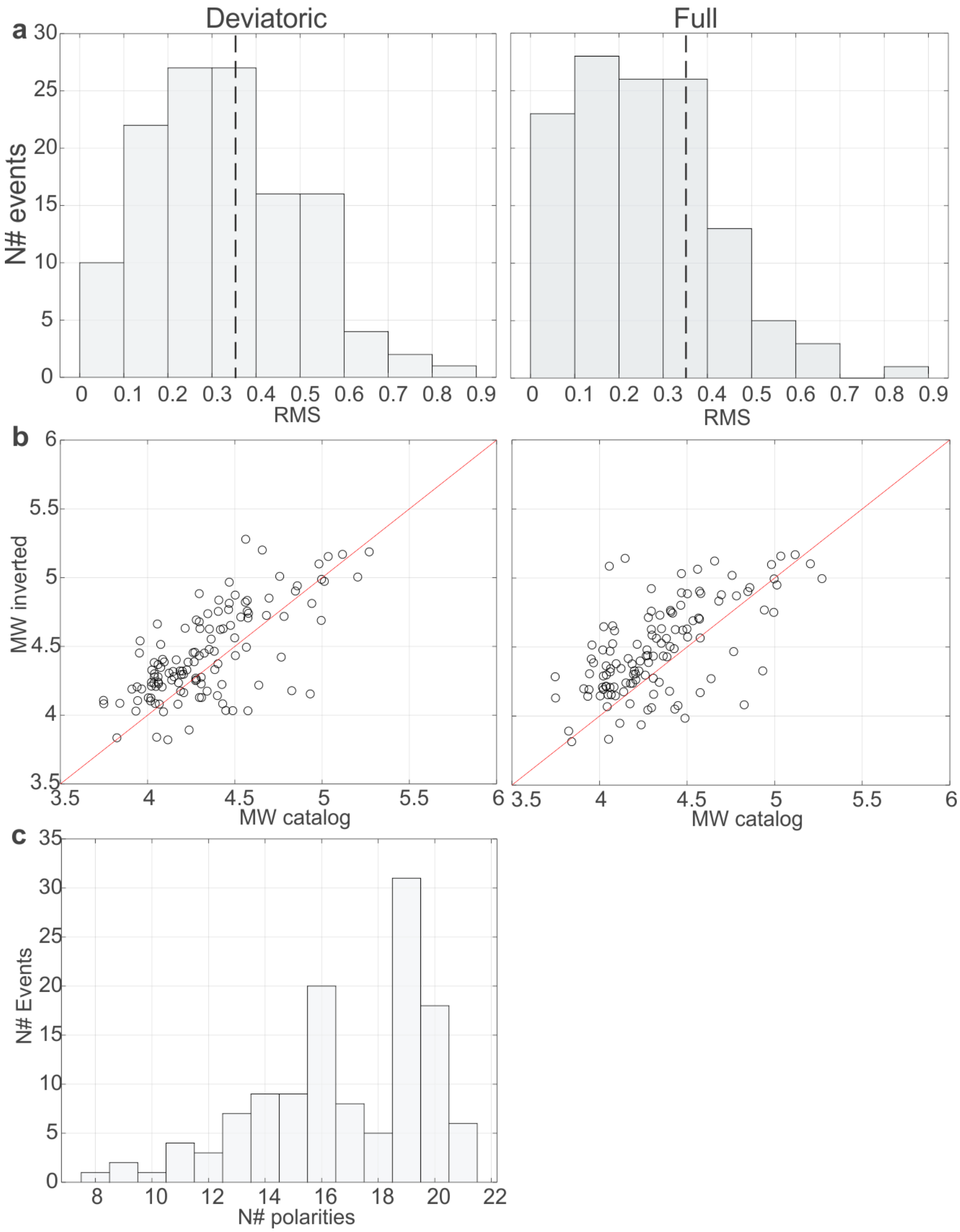

**Figure S7.** Summary of statistics related to the deviatoric (left column) and full (right column) moment tensor inversions. a) RMS distribution for the MT solutions in the deviatoric and full moment tensor inversions respectively. b) Comparison between moment magnitude from NOA catalogue and inverted moment magnitude. Red line indicates the one-to-one. c) Histogram showing the number of stations used in the moment tensors.

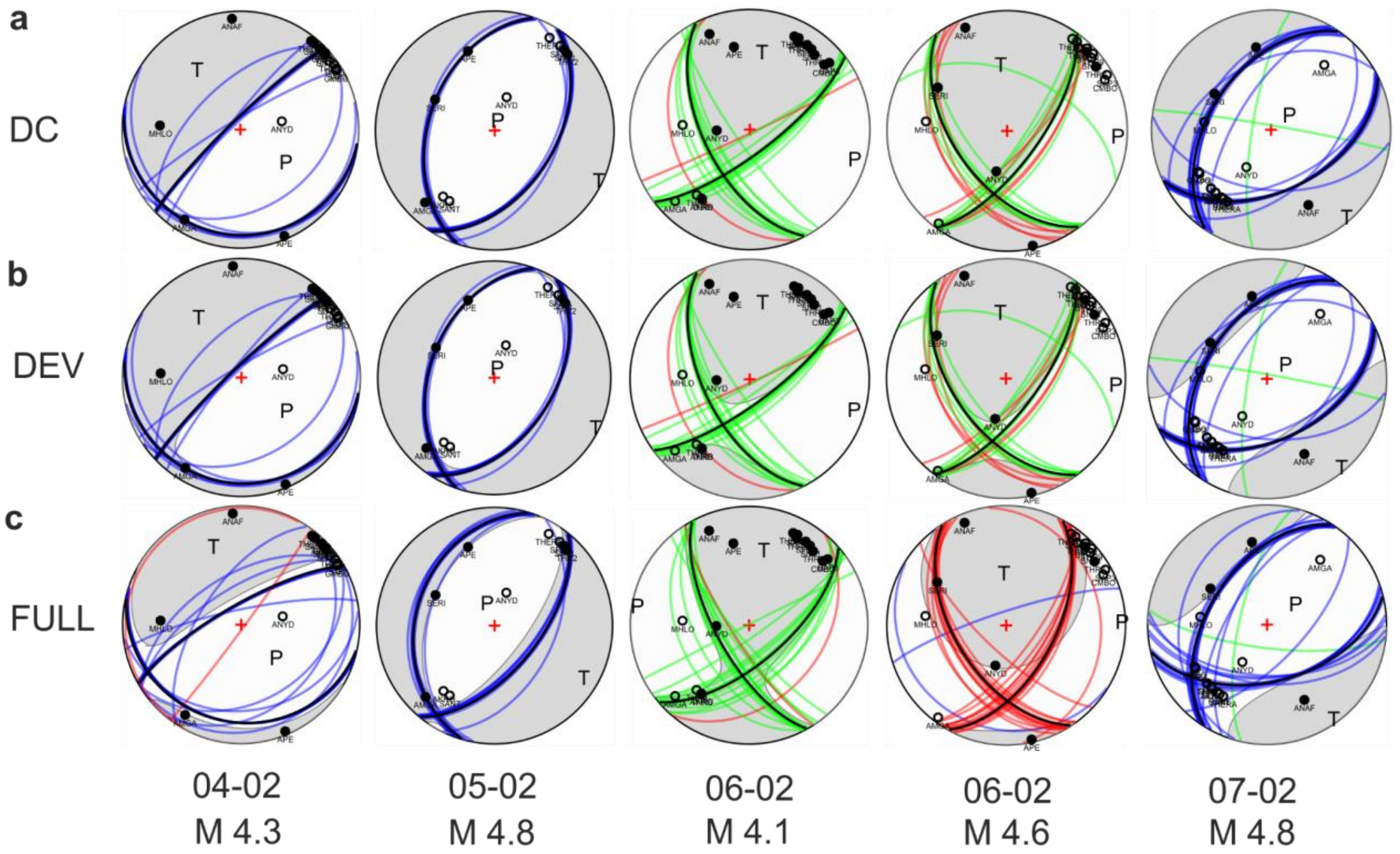

**Figure S8.** Representative examples of moment tensors of M > 4 events during the February 2025 Santorini crisis. Upper, Middle and Bottom rows correspond to solutions from the double-couple constrained (a), deviatoric (b) and full (c) moment tensor inversions, respectively. Columns represent different seismic events, with their date of occurrence and estimated magnitude (NOA) provided at the bottom. The plotted solutions correspond to the best solutions using all the stations, while coloured lines represent the solutions from the Jacknife test. Blue, green, and red colours indicate the obtained faulting styles: normal faulting, strike-slip, and reverse faulting, respectively.

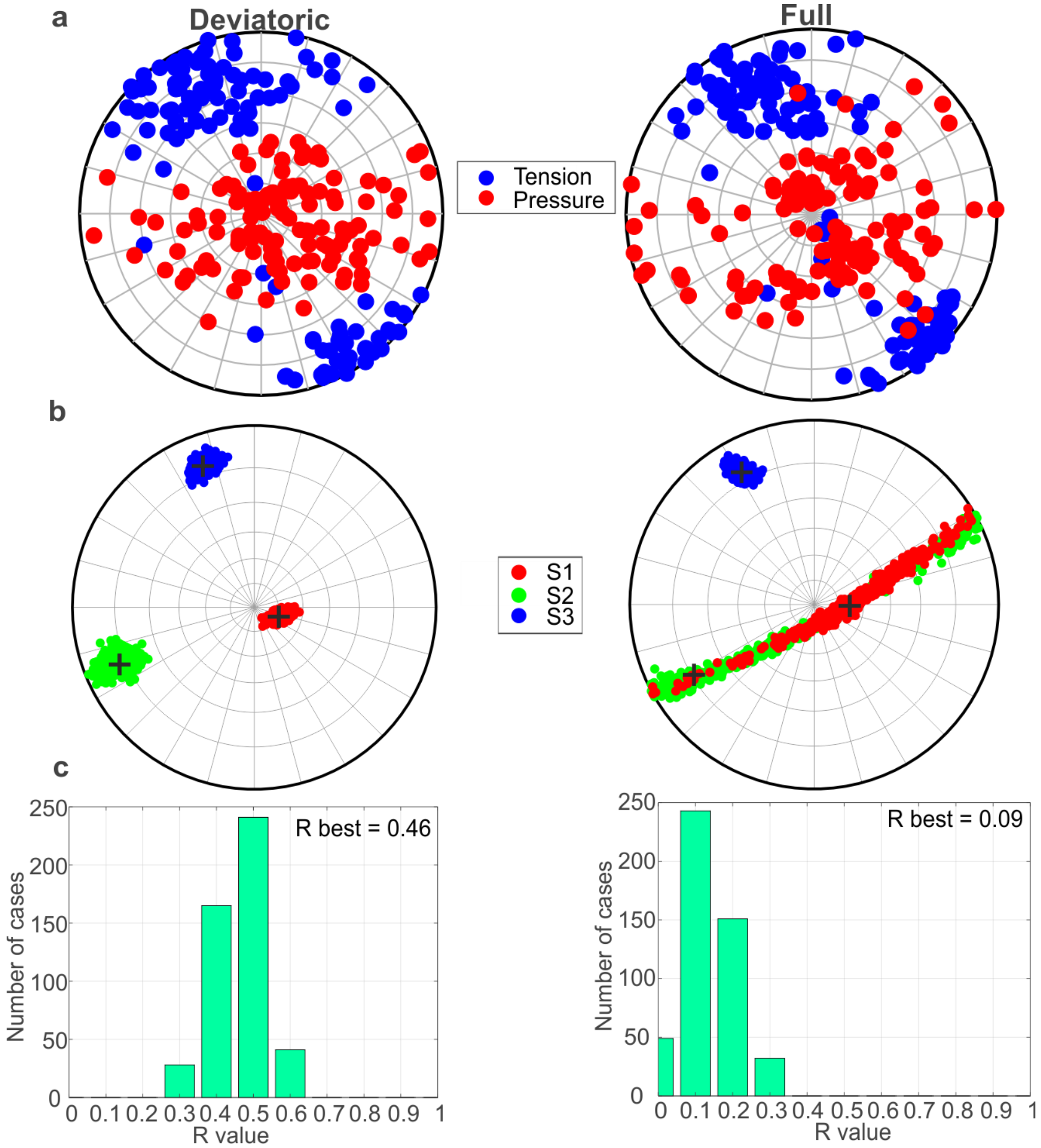

**Figure S9**. Distribution of pressure and tensional axes and insights into the stress field using the fault planes from the deviatoric (left column) and full (right column) moment tensor solutions. a) Distribution of P (red) and T (blue) axis, b) Orientation of the principal stress axis S1 (most compressive, red), S2 (green) and S3 (least compressive, blue).

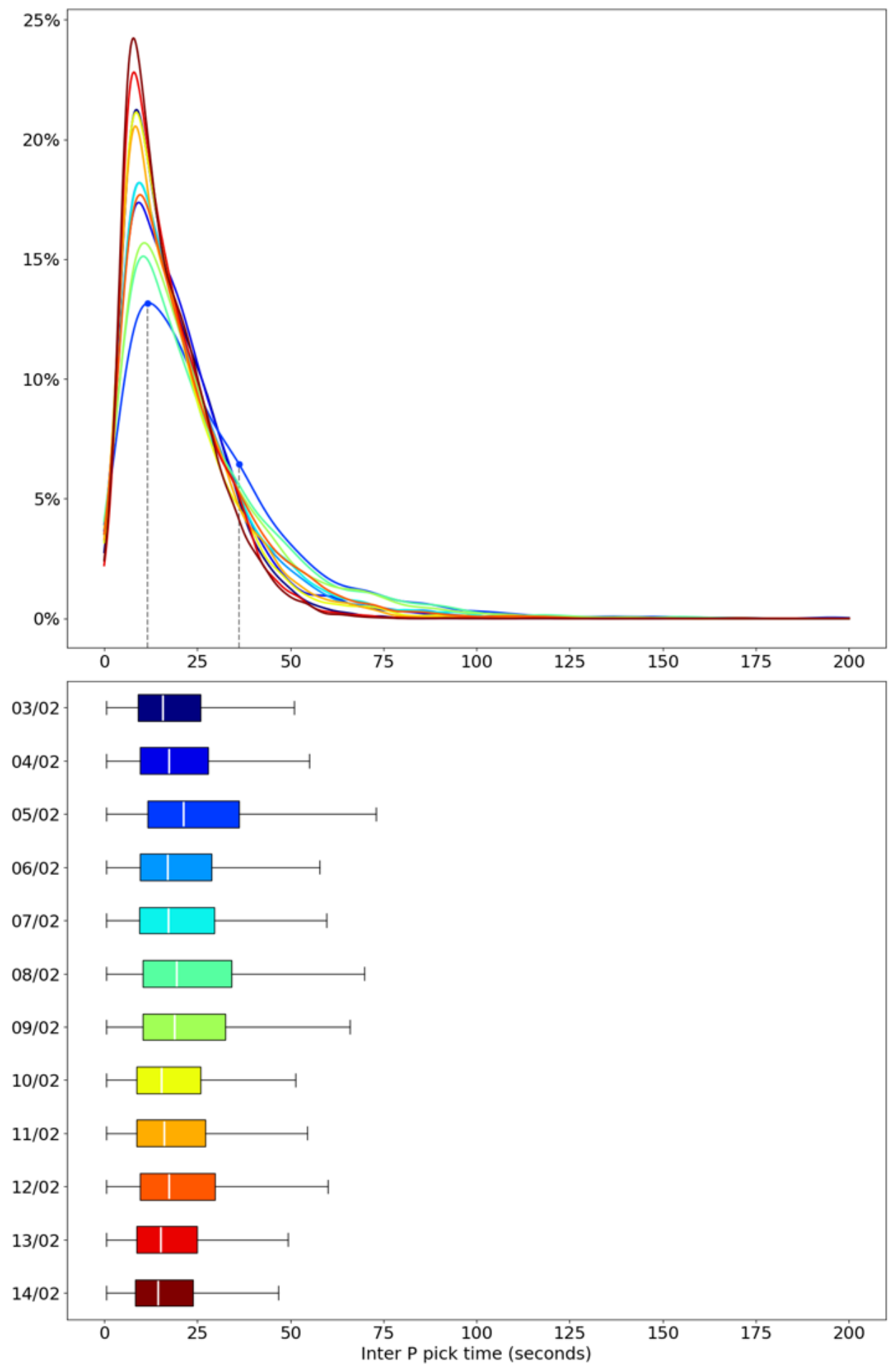

**Figure S10**. Characterising seismicity using phase picks. Top panel: Daily distributions of inter-event times of PhaseNet (Zhu & Beroza, 2019) detections during the increased seismicity period of the seismic unrest, normalised by the total number of P-picks per day. Bottom panel: Width of daily phase pick distributions. The boxes represent the interquartile range ($25^{th}$ to $75^{th}$ percentiles) of the non-parametric distributions. The white vertical lines represent the $50^{th}$ percentiles. Populations of inter-event P-picks with >20% correspond to the early *Onset* (3 February) and *Post-Peak* period (10–11, 13–14 February).

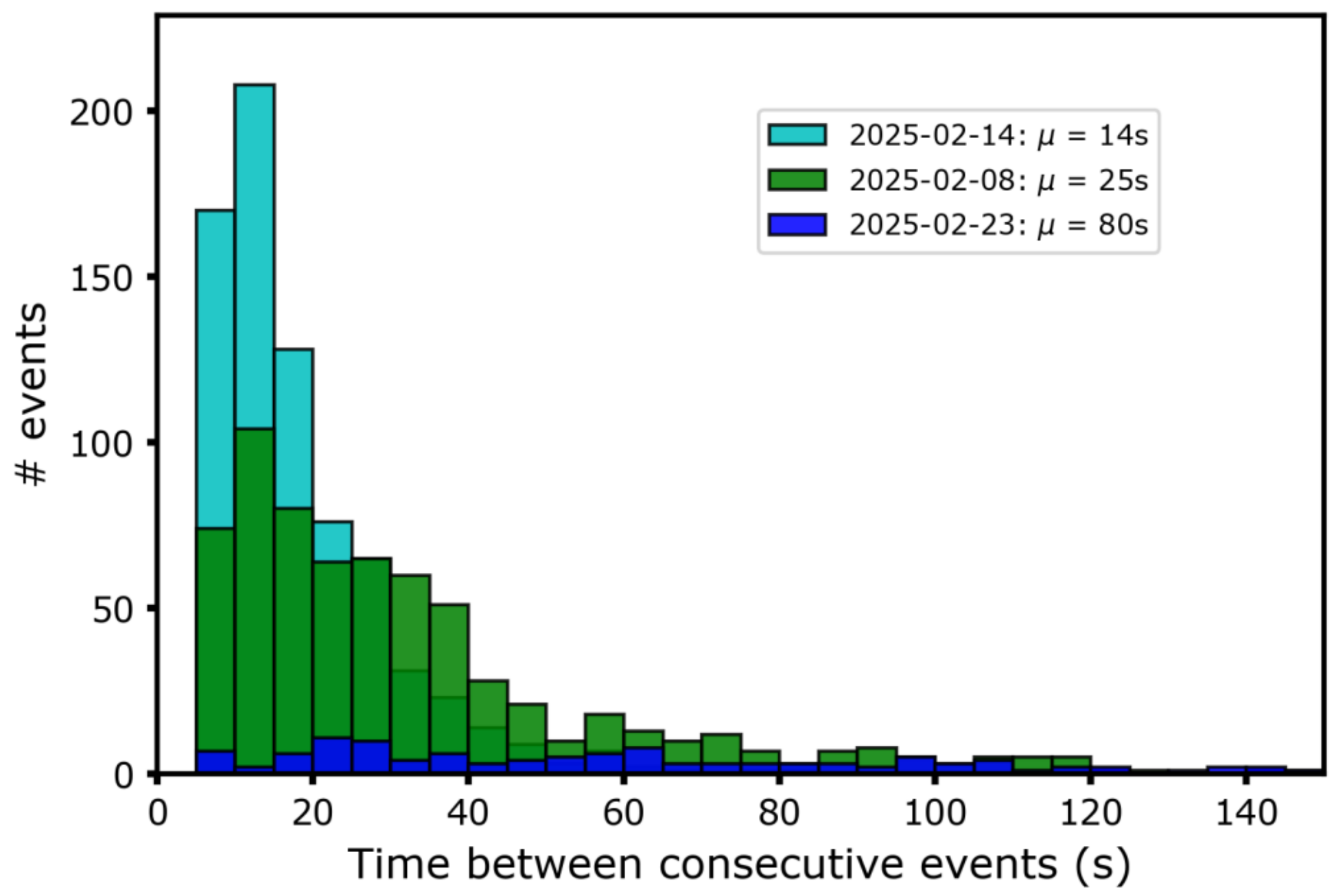

**Figure S11.** Distribution of the inter-event times among consecutive seismic events from the enhanced (by template matching) DL catalogue within 6 hours after M ~ 4 earthquakes in different stages of the seismic crisis. High productivity is observed during the *Peak* phase (2025-02-08) and the "*Is this Volcanic Tremor*?" episode (2025-02-14), while a lower number of earthquakes is observed during the *Deceleration* phase (e.g., 2025-02-23).

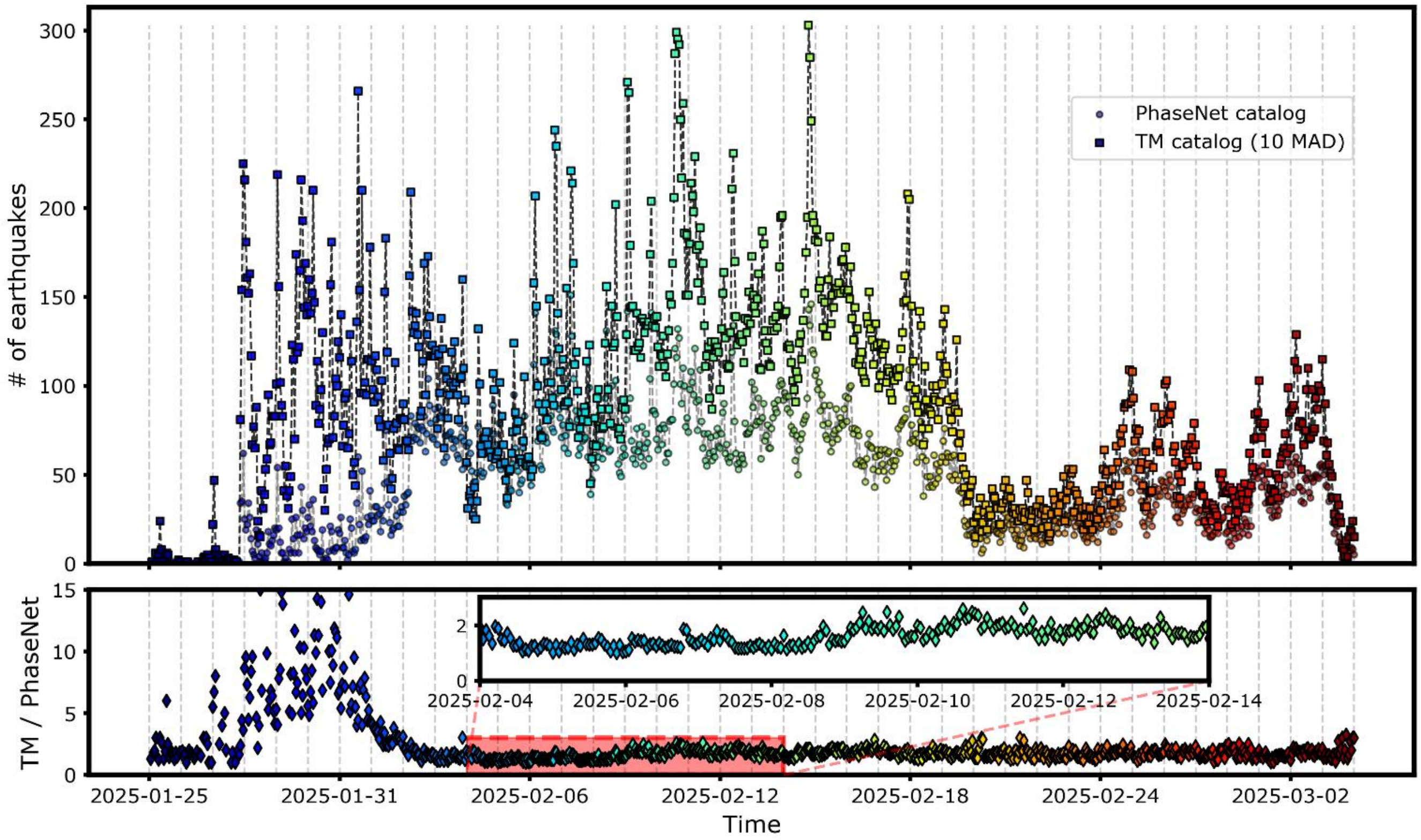

**Figure S12.** Upper panel: Time distribution of the earthquakes identified by the DL and the template matching-enhanced DL catalogue in 1-hour bins. Bottom panel: time distribution of the ratio between the number of earthquakes identified by template matching and machine learning in the 1-hour bins reported in the upper panel. The time distribution highlights a consistent two-fold increase in the number of earthquakes introduced by template matching, with local maxima around the high-seismicity rate period, highlighted as red, described in Figure 5.

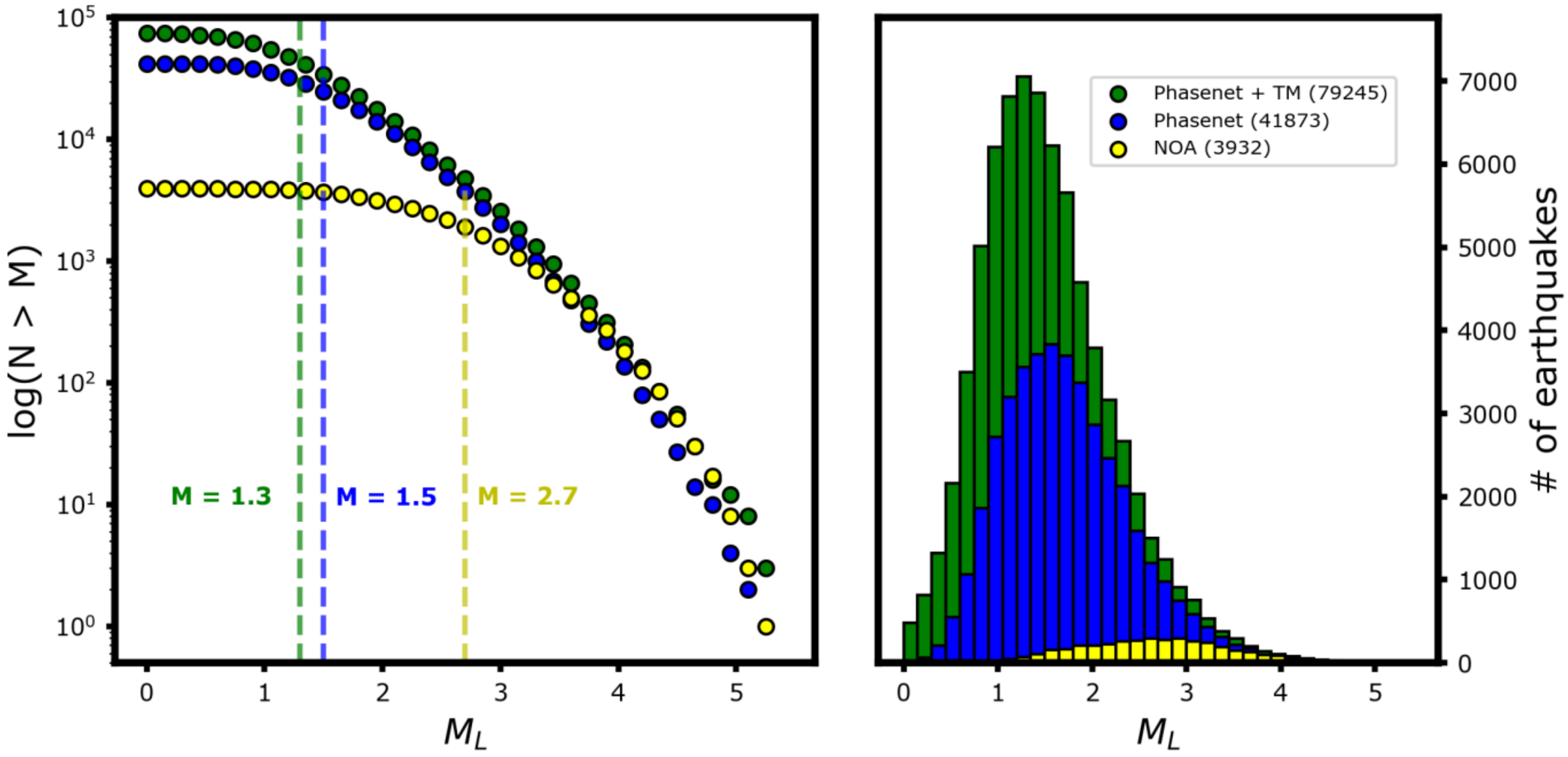

**Figure S13.** Left panel: Gutenberg–Richter distribution for the enhanced DL (green), QF (blue) and NOA (yellow) catalogues during January 25–March 3, 2025 (inclusive). The slope change in the distributions of the enhanced catalogues is more than one magnitude unit lower than the corresponding change for the NOA catalogue. Right panel: linear discrete distribution of the number of earthquakes in the enhanced DL (green), QF (blue) and NOA (yellow) catalogues as a function of magnitude. The use of template matching in the enhanced DL catalogue strongly increases the population of earthquakes with magnitude lower than $M_L$ = 2, nearly doubling the content of the QF catalogue.

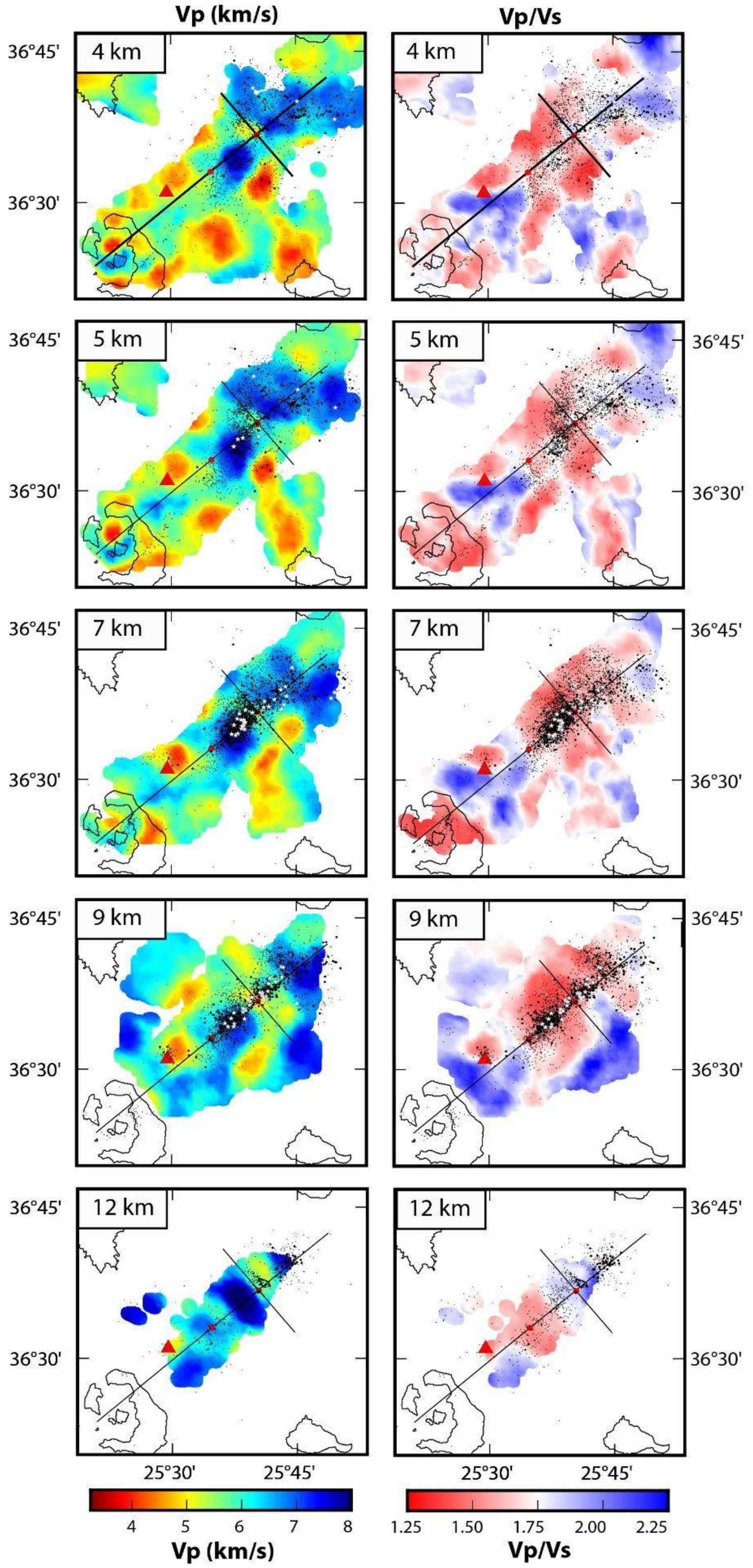

**Figure S14.** Map views at different depths of the 3D $V_p$ and $V_p/V_s$ mean posterior models. Plotted seismicity within 1 km, sized according to local magnitude (small dots: Ml < 3, medium dots: 3 < Ml < 4, stars: Ml > 4).

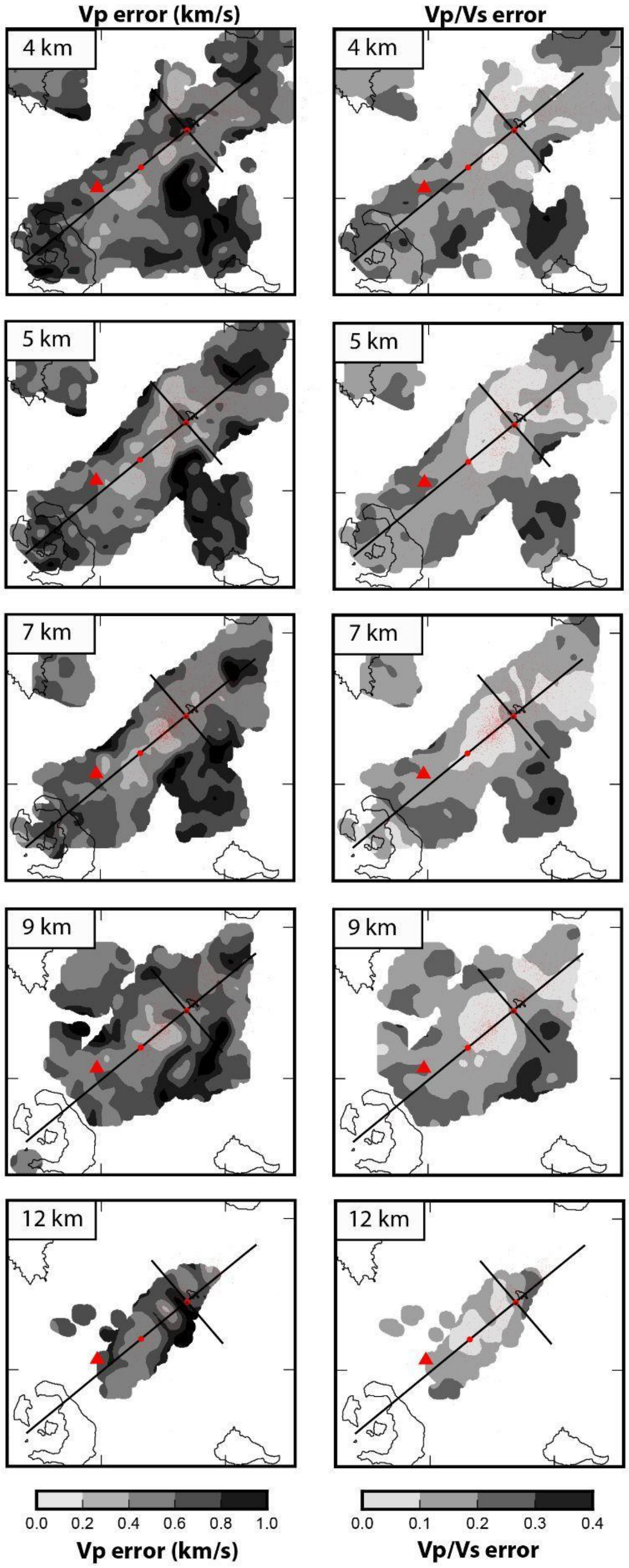

**Figure S15.** Map views at different depths of model errors (absolute deviations), computed from the $V_p$ and $V_p/V_s$ posterior distributions.

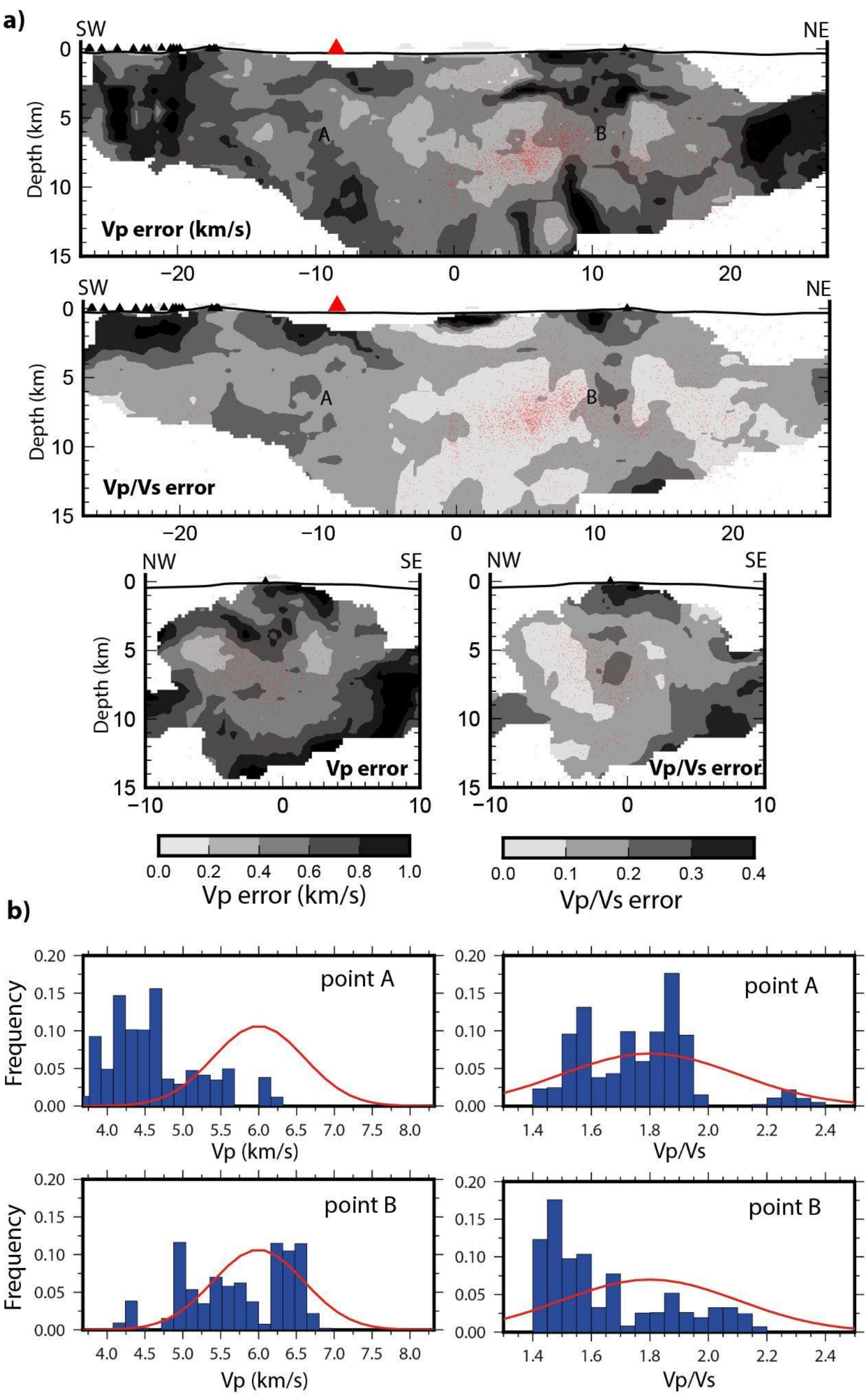

**Figure S16. a)** Vertical cross-sections showing model errors (absolute deviations), computed from the $V_p$ and $V_p/V_s$ posterior distributions (5000 models sampled by 20 different Markov chains starting from a different random point of the model parameter space). **b**) Comparison between prior density distributions (red curve) and 1D marginal posterior distributions (blue bars), computed at two model points (locations in **a**). The multimodal character of the distributions highlights data inconsistency, possibly due to transient velocity changes not captured by the 3D static model.

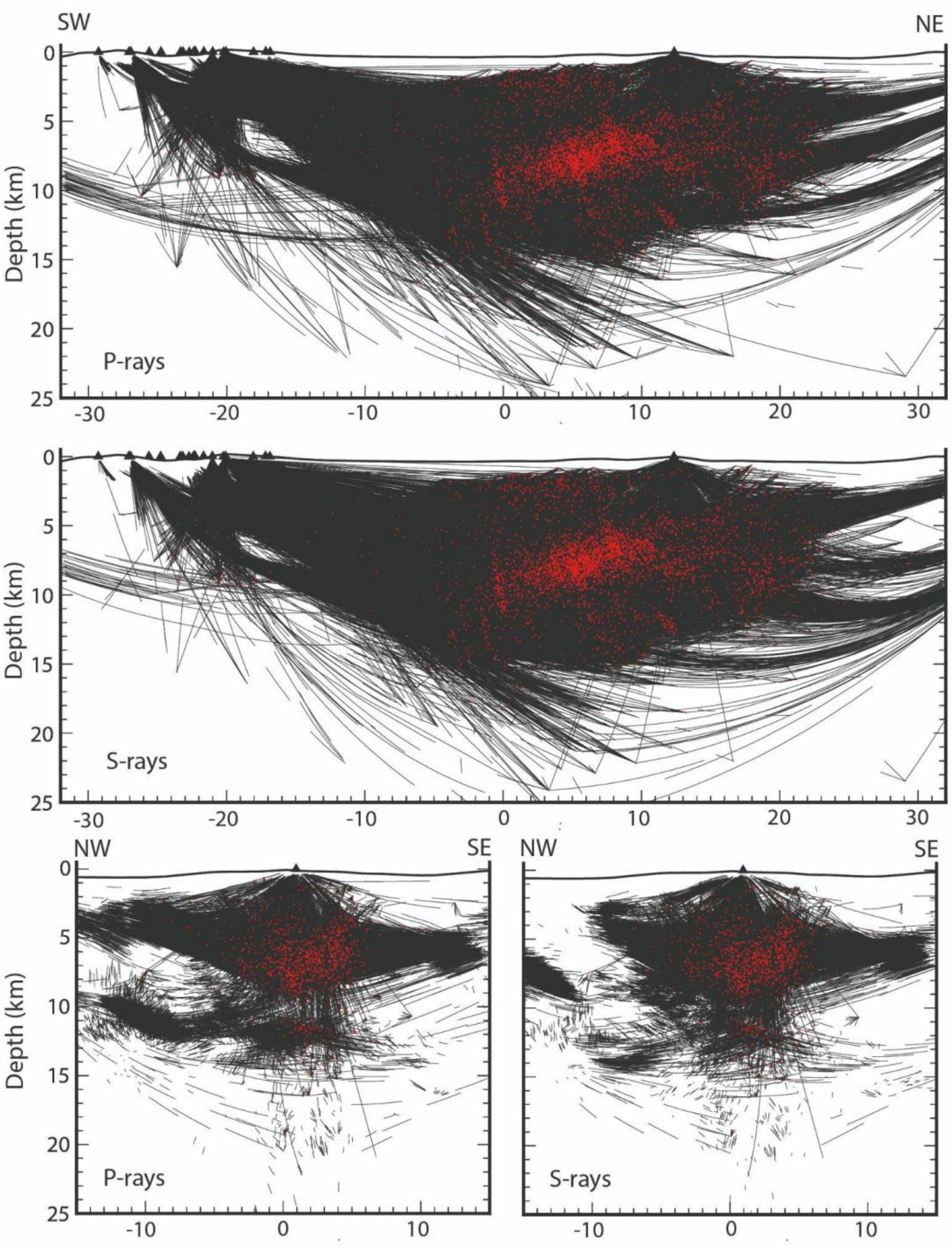

**Figure S17.** Vertical cross-sections showing the coverage of P- and S-wave ray paths. Ray paths and seismicity are projected within 1 km from both sides of the section.

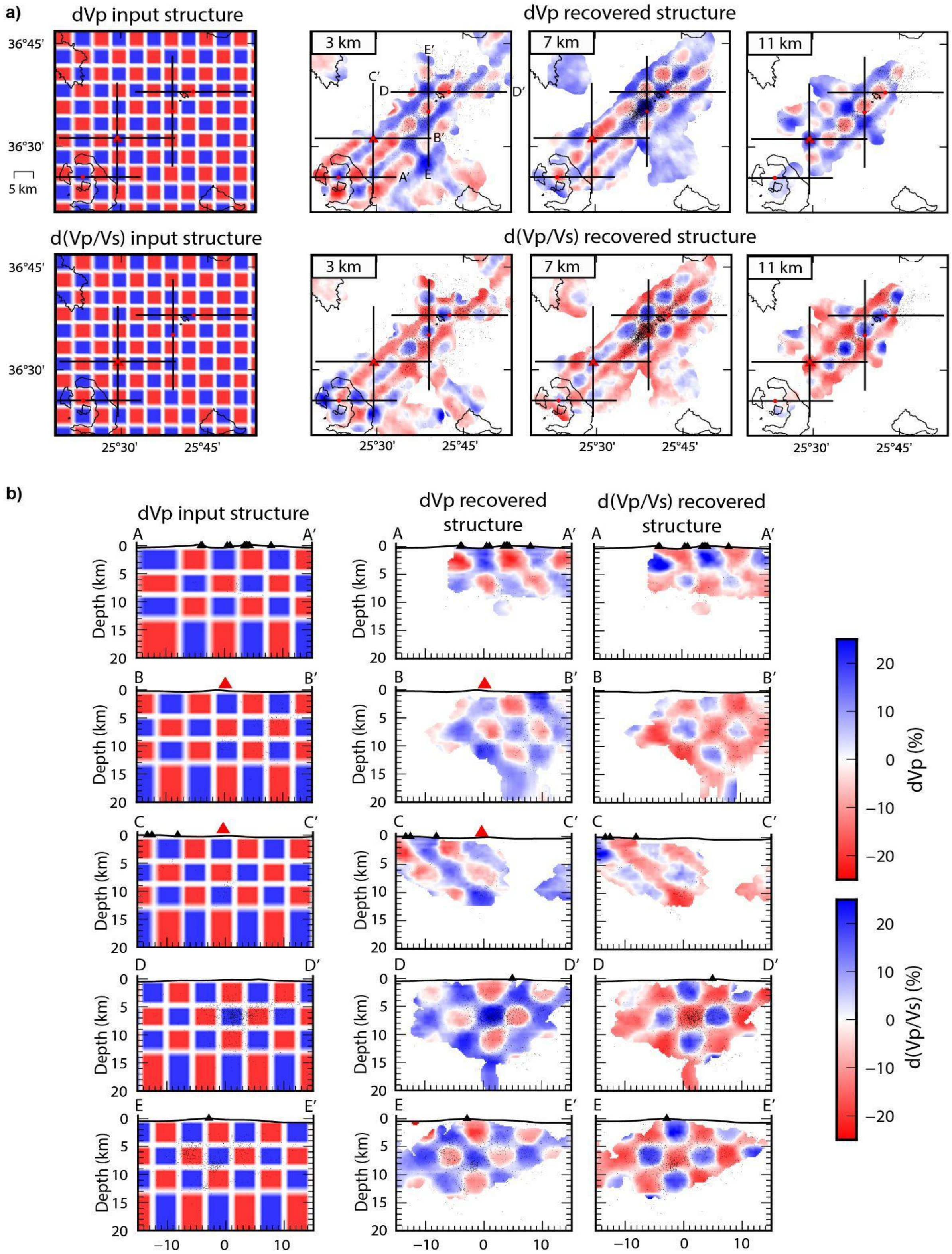

**Figure S18.** Checkerboard resolution test with anomalies of 20% amplitude relative to the background 1D model (of opposite sign between $V_p$ and $V_p/V_s$), and a spatial dimension of 4–5 km in all three directions. Input and recovered models are shown on depth slices (**a**) and vertical cross-sections (**b**).

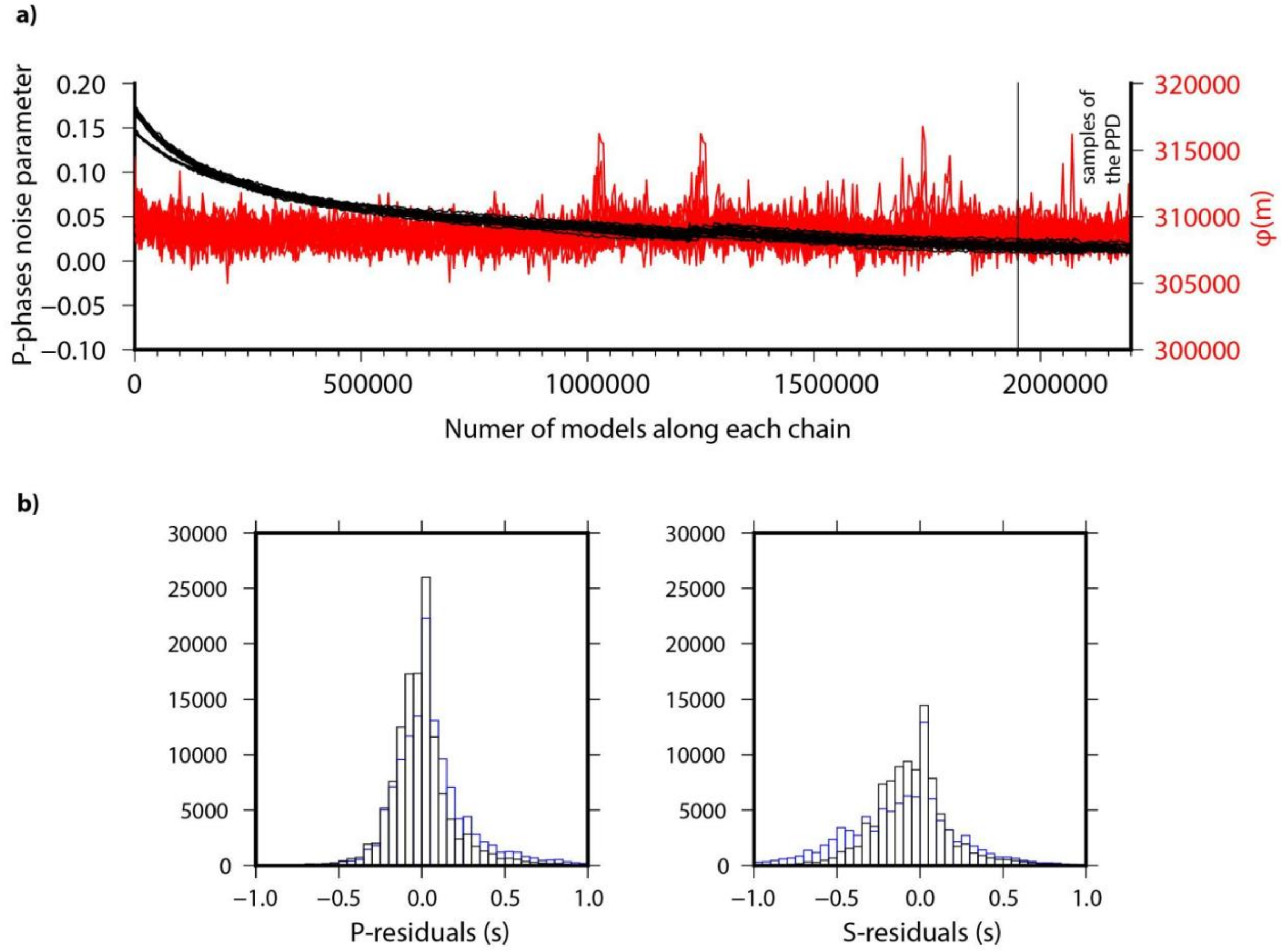

**Figure S19. a**) Evolution of misfit (red lines) and P-phase noise parameter (black lines) along the 20 Markov chains run in the 3D tomography. The vertical line marks the last part of the chain where models are considered samples of the posterior density distribution and thus collected (one every 1000). **b**) Frequency distribution of P-residuals (left panel) and S-residuals (right panel), computed in the prior 1D velocity model (blue) and in the 3D mean posterior velocity model (black).

## Tables

| **Model parameter** | **Prior distribution** | **Min/Mean value** | **Max/$\sigma$ value** |
|---|---|---|---|
| Hypocentral coordinates x, y, z | Normal | GaMMA catalogue | 2.5 km |
| Origin time (ot) | Uniform | -1 s | 1 s |
| Noise level (log scale) | Uniform | -1 | 1 |
| | | | |
| **Model parameter** | **Prior distribution** | **Min/Mean value** | **Max/$\sigma$ value** |
| Number of Voronoi cells | Uniform | 100 | 2000 |
| Position of the Voronoi cells | Uniform | (0, 0, 0)* | (80 km, 80 km, 35 km) |
| $V_p$, $V_p/V_s$ of the Voronoi cells | Normal | 1D model** | 0.5 km/s - 0.3 |
| Hypocentral coordinates x, y, z | Normal | relocation in the 1D model | 1.5 km |
| Origin time (ot) | Uniform | -0.75 s | +0.75 s |
| Noise level (log scale) | Uniform | -1 | 1 |

**Table S1.** Prior distributions of model parameters used in preliminary event relocation (top) and 3D tomography (bottom). Min/Max and Mean/Standard deviation ($\sigma$) refer to the uniform and normal distributions respectively. Normal distributions are sampled in between $\pm 4\sigma$ and uniform distributions in between Min-Max values. *The distribution is uniform in areas crossed by seismic rays. Voronoi cells are not created in regions not illuminated by rays in order to save computation time but can be moved into the null space when their position is perturbed. **(Bohnhoff et al., 2006).

| Weight class | P-phases | S-phases | Starting picking error |
|---|---|---|---|
| 0 | 33933 | 2854 | 0.08 s |
| 1 | 63398 | 29174 | 0.12 s |
| 2 | 31270 | 63922 | 0.2 s |
| 3 | 32521 | 51453 | 0.6 s |

**Table S2.** Distribution of P- and S-wave data used for tomography.

| Model parameters | Probability | Move description |
|---|---|---|
| Hypocentral coordinates x, y, z | 0.9 | Perturb one coordinate of 1 event |
| Noise level | 0.1 | Perturb P-noise (0.05) or S-noise (0.05) |
| | | |
| **Model parameters** | **Probability** | **Move description** |
| Number of Voronoi cells | 0.15 | Create or destroy up to 2 cells |
| Position of Voronoi cells | 0.15 | Perturb coordinates of 2 Voronoi cells |
| $V_p$ and $V_p/V_s$ of Voronoi cells | 0.2 | Perturb the $V_p$ and $V_p/V_s$ of 2 Voronoi cells |
| Hypocentral coordinates x, y, z | 0.4 | Perturb one coordinate of 2 events |
| Noise level | 0.1 | Perturb P-noise (0.05) or S-noise (0.05) |

**Table S3.** Recipe of model sampling along Markov chains used in preliminary event relocation (top) and 3D tomography (bottom).

## Acknowledgements

The work of BGS scientists has been supported by the AI in Seismology International Geoscience Research and Development (IGRD) project. The research was part of the BGS International NC programme 'Geoscience to tackle Global Environmental Challenges', NERC reference NE/X006255/1. The paper is published by permission of the Director of the British Geological Survey. BGS advisories related to the Santorini unrest, submitted to the Foreign, Commonwealth and Development Office (FCDO), were informed by our results. We thank Sotiris Valkaniotis for providing geological data in vector format. The GNSS data are made available by Greek (Hellenic Positioning System HEPOS, NOA GNSS Network NOANET, National Technical University of Athens, National and Kapodistrian University of Athens, University of Patras, Hexagon SmartNet provider, URANUS GNSS Network, JGC Geoinformation Systems S.A.) and international (CNRS, INGV) GNSS networks under the coordination of the Greek Earthquake Planning and Protection Organisation (EPPO). Any use of trade, firm, or product names is for descriptive purposes only and does not imply endorsement by the U.S. Government.

## Data availability

Waveforms were accessed daily from the NOA EIDA open data node (https://eida.gein.noa.gr) (Evangelidis et al., 2021). The data used in the search for global analogues are available at https://earthquake.usgs.gov/earthquakes/search/ (ComCat), https://www.isc.ac.uk/ (ISC) (International Seismological Centre, 2026), http://hraun.vedur.is/ja/viku/ (SIL) (Bodvarsson et al., 1996), https://www.jma.go.jp/jma/index.html (JMA), https://www.ign.es/web/ign/portal/sis-catalogo-terremotos (IGN) and https://volcano.si.edu/ (GVP) (Global Volcanism Program & Venzke, 2025). The earthquake catalogues produced in this study are available on Zenodo (Segou et al., 2026).